\documentclass[onecolumn,aps,superscriptaddress,nofootinbib,preprint, prd]{revtex4-2}
\usepackage[T1]{fontenc}

\usepackage{amsmath,amssymb,amsfonts,bm,slashed,graphicx,longtable,xcolor,multirow,xspace}
\usepackage{array}[=2016-10-06]
\usepackage{multirow}
\usepackage{breakurl}
\usepackage[shortlabels]{enumitem}
\usepackage{placeins}
\usepackage{mciteplus}
\usepackage{graphicx,tikz,pgflibraryshapes,xcolor}
\usetikzlibrary{decorations,trees}
\usetikzlibrary{positioning,shapes}
\usetikzlibrary{decorations.text}
\tikzstyle{every picture}+=[remember picture]
\pgfdeclarelayer{background}
\pgfdeclarelayer{foreground}
\pgfsetlayers{background,main,foreground}

\usepackage{xparse}
\tikzset{
	circleps/.style={circle, path picture={
	\begin{pgfonlayer}{background}
  	\draw[black, line width = 2pt]	(path picture bounding box.south east) -- (path picture bounding box.north west) (path picture bounding box.south west) -- (path picture bounding box.north east);
	\end{pgfonlayer}}}
}
\DeclareDocumentCommand{\cvertex}{O{}}{\tikz{\node [anchor=base, circle#1, fill=black!80, draw = black!80, line width = 1pt, inner sep = 3pt]  {};}}
\DeclareDocumentCommand{\overtex}{O{}}{\tikz{\node [anchor=base, circle#1, fill=white, draw = black!80, line width = 1pt, inner sep = 3pt]  {};}}
\DeclareDocumentCommand{\puredge}{O{} O{}}{\tikz{
	\node (l) [anchor=base,circle#1, fill=white, draw = black!80, line width = 1pt, inner sep = 3pt]  {};
	\draw (l) node (r) [anchor=base,circle#2, fill=white, draw = black!80, line width = 1pt, inner sep = 3pt, xshift = 20pt]  {};
	\draw [double, draw = black!80, line width = 2pt] (l) -- (r);
	}}
\DeclareDocumentCommand{\paredge}{O{} O{}}{\tikz{
	\node (l) [anchor=base,circle#1, fill=black!80, draw = black!80, line width = 1pt, inner sep = 3pt]  {};
	\draw (l) node (r) [anchor=base,circle#2, fill=white, draw = black!80, line width = 1pt, inner sep = 3pt, xshift = 20pt]  {};
	\draw [double, draw = black!80, line width = 2pt] (l) -- (r);
	}}
\DeclareDocumentCommand{\fuledge}{O{} O{}}{\tikz{
	\node (l) [anchor=base,circle#1, fill=black!80, draw = black!80, line width = 1pt, inner sep = 3pt]  {};
	\draw (l) node (r) [anchor=base,circle#2, fill=black!80, draw = black!80, line width = 1pt, inner sep = 3pt, xshift = 20pt]  {};
	\draw [double, draw = black!80, line width = 2pt] (l) -- (r);
	}}
\newcommand{\vvertex}[1]{\tikz{\node [anchor=base,circle, fill=white, draw = black!80, line width = 1pt, inner sep = 1pt]  {#1};}}
\newcommand{\vedge}[2]{\tikz{
	\node (l) [circle, fill=white, draw = black!80, line width = 1pt, inner sep = 1pt]  {#1};
	\draw (l) node (r) [circle, fill=white, draw = black!80, line width = 1pt, inner sep = 1pt, xshift = 30pt]  {#2};
	\draw [double, draw = black!80, line width = 2pt] (l) -- (r);
	}}

\usepackage{listings,textcomp}
\lstnewenvironment{activett}{}{}

\definecolor{solarized@base03}{HTML}{002B36}
\definecolor{solarized@base02}{HTML}{073642}
\definecolor{solarized@base01}{HTML}{586e75}
\definecolor{solarized@base00}{HTML}{657b83}
\definecolor{solarized@base0}{HTML}{839496}
\definecolor{solarized@base1}{HTML}{93a1a1}
\definecolor{solarized@base2}{HTML}{EEE8D5}
\definecolor{solarized@base3}{HTML}{FDF6E3}
\definecolor{solarized@yellow}{HTML}{B58900}
\definecolor{solarized@orange}{HTML}{CB4B16}
\definecolor{solarized@red}{HTML}{DC322F}
\definecolor{solarized@magenta}{HTML}{D33682}
\definecolor{solarized@violet}{HTML}{6C71C4}
\definecolor{solarized@blue}{HTML}{268BD2}
\definecolor{solarized@cyan}{HTML}{2AA198}
\definecolor{solarized@green}{HTML}{859900}
\definecolor{darkred}{HTML}{550003}
\definecolor{darkgreen}{HTML}{00AA00}

\newcommand\CPPidentifierstyle{\color{solarized@blue}\footnotesize\ttfamily}
\newcommand\CPPcommentstyle{\color{solarized@violet}\footnotesize\ttfamily}
\newcommand\CPPdirectivestyle{\color{solarized@magenta}\footnotesize\ttfamily}

\lstdefinestyle{cpp}
{
  language=C++,
  basicstyle=\footnotesize\ttfamily,
  basewidth={0.53em,0.44em}, 
  numbers=none,
  tabsize=2,
  breaklines=true,
  escapeinside={(@}{@)}, 
  showstringspaces=false,
  numberstyle=\tiny\color{solarized@base01},
  keywordstyle=\color{solarized@orange},
  stringstyle=\color{solarized@red}\ttfamily,
  identifierstyle=\color{solarized@blue},
  commentstyle=\CPPcommentstyle,
  directivestyle=\CPPdirectivestyle,
  emphstyle=\color{solarized@green},
  frame=single,
  rulecolor=\color{solarized@base2},
  rulesepcolor=\color{solarized@base2},
  literate={~} {\customtilde}1,
  moredelim=*[directive]\ \ \#,
  moredelim=*[directive]\ \ \ \ \#,
  }
\newcommand\textlst[1]{\lstinline!#1!}

\DeclareDocumentCommand{\cmark}{O{}}{\tikz{\node [anchor=base, #1, fill=black!50!white, draw = none, line width = 1pt, inner sep = 3pt]  {};}}

\newcommand{\hammer}{\texttt{Hammer}\xspace}

\newcommand{\nn}{\nonumber}

\newcommand{\mn}{{\mu\nu}}
\newcommand{\g}{\gamma}
\newcommand{\ampB}[2]{\big\langle #1 \big|\, #2\, \big| B \big\rangle }
\newcommand{\ampBb}[2]{\big\langle #1 \big|\, #2\, \big| \Bbar \big\rangle }

\newcommand{\dotpr}[2]{\,#1\!\cdot #2\, }

\newcommand{\PS}{\mathcal{PS}}
\def\PSp{\PS}
\newcommand{\BR}{\mathcal{B}}

\newcommand{\dds}{D^{(*)}}
\newcommand{\dss}{D^{**}}
\newcommand{\ddss}{D^{(*,**)}}

\newcommand{\dSs}{D^*_0}
\newcommand{\dVs}{D^*_1}
\newcommand{\dVp}{D_1}
\newcommand{\dTs}{D^*_2}
\newcommand{\bbar}{\bar{b}}
\newcommand{\cbar}{\bar{c}}
\newcommand{\ubar}{\bar{u}}

\newcommand{\Bbar}{\,\overline{\!B}{}}

\newcommand{\lqcd}{\ensuremath{\Lambda_{\rm QCD}}\xspace}

\newcommand{\Lb}{\Lambda_b\xspace}
\newcommand{\Lc}{\Lambda_c\xspace}

\newcommand{\LbLcsln}{{\Lb \to \Lc^{(*)} \ell \nu}}

\newcommand{\tC}{\chi}
\newcommand{\tL}{\lambda}
\newcommand{\chSL}{\tC_L^S}
\newcommand{\chSR}{\tC_R^S}
\newcommand{\chVL}{\tC_L^V}
\newcommand{\chVR}{\tC_R^V}
\newcommand{\chTL}{\tC_L^T}
\newcommand{\chTR}{\tC_R^T}
\newcommand{\laSL}{\tL_L^S}
\newcommand{\laSR}{\tL_R^S}
\newcommand{\laVL}{\tL_L^V}
\newcommand{\laVR}{\tL_R^V}
\newcommand{\laTL}{\tL_L^T}
\newcommand{\laTR}{\tL_R^T}

\newcommand{\alSL}{\alpha_L^S}
\newcommand{\alSR}{\alpha_R^S}
\newcommand{\alVL}{\alpha_L^V}
\newcommand{\alVR}{\alpha_R^V}
\newcommand{\alTL}{\alpha_L^T}
\newcommand{\alTR}{\alpha_R^T}

\allowdisplaybreaks

\makeatletter
\g@addto@macro\bfseries{\boldmath}
\makeatother

\graphicspath{{./figures/}}

\definecolor{red}{rgb}{0.9, 0,0}

\makeatletter
\let\tmp@footnote\footnote
\renewcommand{\footnote}[1]{\tmp@footnote{\linespread{0.9}\selectfont{}#1}}
\makeatother

\DeclareRobustCommand{\SkipTocEntry}[5]{}

\usepackage[colorlinks,citecolor=blue,urlcolor=blue,linkcolor=blue,breaklinks=breakall]{hyperref}

\begin{document}
\preprint{CALT-TH-2026-027}

\title{An introduction to Hammer v2: \\ Helicity Amplitude Module for Matrix Element Reweighting\\ \small{ver. 2.0.0}}

\author{Michele Papucci}
\affiliation{Walter Burke Institute for Theoretical Physics
and Leinweber Forum for Theoretical Physics, 
California Institute of Technology, Pasadena, CA 91125, USA}

\author{Dean J.\ Robinson}
\affiliation{Ernest Orlando Lawrence Berkeley National Laboratory
and Leinweber Institute for Theoretical Physics at UC Berkeley,
University of California, Berkeley, CA 94720, USA}

\begin{abstract}
The \texttt{Hammer} software library provides fast and efficient reweighting of large simulated datasets 
containing semileptonic $b$-hadron decays to any beyond Standard Model (BSM) theory, 
or to any form-factor description of the hadronic matrix elements.  
By enabling reweighting to a different underlying theoretical model after the computationally-expensive detector simulation step has already been completed, 
\texttt{Hammer} permits experimental analyses to employ forward-folding fitting strategies 
to recover underlying physical parameters without biases, or to properly characterize theory systematic uncertainties. 
This publication details upgrades to \texttt{Hammer} functionalities and its application programming interface (API) for version \texttt{2.x}, 
and also provides associated documentation of the library's structure, syntactical conventions, and code flow.
Substantial optimization of \texttt{Hammer}'s internal tensor library now enables computational complexity to generically scale almost linearly with amplitude tensor rank 
times size rather than as a quartic,
enabling reweighting into very high dimension spaces, 
such as the product of BSM Wilson coefficient and form factor parameter linear spaces, while also retaining Monte Carlo uncertainties. 
Updated \texttt{Python} bindings are implemented with bijective correspondence to the \texttt{C++} interface, allowing full access to library functionalities.
\end{abstract}

\makeatletter
\let\temp@clearpage\clearpage
\let\clearpage\relax
\makeatother

\maketitle

\addtocontents{toc}{\SkipTocEntry}
\section*{Hammer collaboration}

\scalebox{0.85}{
\begin{minipage}{\textwidth}
	Florian U.~Bernlochner\footnotemark[1], Stephan Duell\footnotemark[1], Zoltan Ligeti\footnotemark[2]\,\footnotemark[3],\\ 
	Michele Papucci\footnotemark[4]\,\footnotemark[5], Dean~J.~Robinson\footnotemark[2]\,\footnotemark[3]\,\footnotemark[5] \\[8pt]
	Webpage:~\href{https://hammer.physics.lbl.gov}{hammer.physics.lbl.gov}  \quad Source code:~\href{https://gitlab.com/mpapucci/Hammer}{git repository}.
	
	\footnotetext[1]{Physikalisches Institut der Rheinischen Friedrich-Wilhelms-Universit\"at Bonn, 53115 Bonn, Germany}
	\footnotetext[2]{Ernest Orlando Lawrence Berkeley National Laboratory, \\[-5pt]
University of California, Berkeley, CA 94720, USA}
	\footnotetext[3]{Leinweber Institute for Theoretical Physics at UC Berkeley, \\[-5pt]
University of California, Berkeley, CA 94720, USA}
	\footnotetext[4]{Walter Burke Institute for Theoretical Physics and Leinweber Forum for Theoretical Physics, \\[-5pt]
California Institute of Technology, Pasadena, CA 91125, USA}
	\footnotetext[5]{Corresponding developer. For support: hammer-support@lbl.gov}
\end{minipage}
}

\makeatletter
\let\clearpage\temp@clearpage
\makeatother

\newpage
\twocolumngrid
 {
 \fontsize{10}{8}\selectfont 
 \columnsep20pt
 	\tableofcontents
 }
\onecolumngrid

\addtocontents{toc}{\SkipTocEntry}
\section*{Version history and DOIs}
\scalebox{0.85}{
\begin{minipage}{\textwidth}
The Hammer collaboration kindly asks that usage of \hammer is accompanied by citations 
to the original release paper~\cite{Bernlochner:2020tfi} and this publication (for version \texttt{2.x}).
In addition, to ensure the future provenance of analyses, we recommend citation of the Zenodo DOI associated with the particular \hammer version that was used, 
per the table below 
(for the latest patch release of every \texttt{major}.\ \texttt{minor} version).
The DOI labelled `All' resolves to the generic package, inclusive of all versions.

\begin{center}
\begin{tabular}{m{2cm}m{5cm}c}
\hline\hline
Version & \quad DOI & Reference \\
\hline
All & \href{https://doi.org/10.5281/zenodo.3722681}{\textlst{10.5281/zenodo.3722681}} \\ 
\hline
2.0.0 & \href{https://doi.org/10.5281/zenodo.20716678}{\textlst{10.5281/zenodo.20716678}} & \cite{bernlochner_2026_20716678}\\
1.4.1 & \href{https://doi.org/10.5281/zenodo.11245573}{\textlst{10.5281/zenodo.11245573}} & \cite{bernlochner_2024_11245573} \\
1.3.0 & \href{https://doi.org/10.5281/zenodo.7007837}{\textlst{10.5281/zenodo.7007837}} & \cite{bernlochner_2022_7007837} \\
1.2.1 & \href{https://doi.org/10.5281/zenodo.5828435}{\textlst{10.5281/zenodo.5828435}} & \cite{bernlochner_2022_5828435} \\
1.1.0 & \href{https://doi.org/10.5281/zenodo.3993770}{\textlst{10.5281/zenodo.3993770}} & \cite{bernlochner_2020_3993770} \\
1.0.0 & \href{https://doi.org/10.5281/zenodo.3722682}{\textlst{10.5281/zenodo.3722682}} & \cite{bernlochner_2020_3722682} \\
\hline
\end{tabular}
\end{center}

\end{minipage}
}

\section{Introduction}
Precision analyses of semileptonic $b$-hadron decays typically rely on detailed numerical Monte Carlo (MC) simulations of detector responses and acceptances.
Combined with the underlying theoretical models, these simulations provide MC \emph{templates} that may be used in fits, to
translate experimental yields into theoretically well-defined parameters.
This translation though can become sensitive to the template and its underlying theoretical model, 
introducing biases whenever there is a mismatch between the theoretical assumptions used to measure a parameter and subsequent theoretical interpretations of the data.

Such biases are known to arise in the analyses of semileptonic decays of $b$ hadrons~\cite{Bernlochner:2020tfi}, 
in particular, for the measurements of the CKM elements $|V_{cb}|$ and $|V_{ub}|$, 
and the lepton flavor universality (LFUV) ratios
\begin{equation}
\label{RMdef}
	R(H_c) = \frac{\Gamma(H_b\to H_c\tau\bar\nu)}{\Gamma(H_b\to H_c \ell \bar\nu)}\,, 
  \qquad \ell = \mu, \,e\,,
\end{equation}
where $H_{b,c}$ denote $b$-~and $c$-flavor hadrons.
To avoid this, the size of these biases need to be either carefully controlled when experiments quote their results by reversing detector effects, 
or they can be avoided by using dedicated MC samples for each theoretical model the measurement is confronted with. 
This manual presents a detailed overview of the capabilities and application programming interface of \hammer\ (\emph{Helicity Amplitude Module for Matrix Element Reweighting}),
designed expressly to enable the latter approach within experimental simulation frameworks via reweighting.

Semitauonic $b$-hadron decays have long been known to be sensitive to new physics~\cite{Krawczyk:1987zj, 
Heiliger:1989yp, Kalinowski:1990ba, Grzadkowski:1991kb, Grossman:1994ax, Tanaka:1994ay, Goldberger:1999yh}, 
and were first constrained at LEP~\cite{Buskulic:1992um}.
Precision analyses of semileptonic $b$-hadron decays also provide the predominant avenue for precision exclusive measurements of $|V_{cb}|$ and $|V_{ub}|$.
At present, the measurements of the $R(\dds)$ ratios show about a $3\sigma$ tension with SM predictions, 
when the $D$ and $D^*$ modes are combined~\cite{HeavyFlavorAveragingGroupHFLAV:2024ctg}. 
In the future, much more precise measurements of semitauonic decays are expected, not only for the $B\to D^{(*)}l\bar\nu$ channels, 
but also for $\Lambda_b\to \Lambda_c l\bar\nu$, $B_s\to D_s^{(*)}l\bar\nu$, $B_c \to J/\psi\, l\bar\nu$,
as well as those involving excited charm hadrons in the final state.
Similarly, exclusive measurements of $|V_{cb}|$ now feature precision at the $2\%$, 
in $\sim 3\sigma$ tension with inclusive measurements~\cite{ParticleDataGroup:2026aaa}.
The exclusive precision may be improvable to the $<1\%$ level with full HL-LHC and Belle~II datasets,
but such precision requires exquisite control over systematic uncertainties, including those of theoretical origin.

Semileptonic analyses rely heavily on large MC simulations to optimize selections, provide fit templates in discriminating kinematic observables, 
and to model resolution effects and acceptances. 
Both the $\tau$ (in semitauonic decays) and the charm hadrons have short lifetimes and decay near the production vertex, 
so that most of the measurements rely on reconstruction of the ensuing decay cascades. 
To reconstruct the decay products, often complex phase space cuts and detector efficiency dependencies are employed: 
the measurement of the full decay kinematics (without at least multifold ambiguities) is impossible due to the presence of multiple neutrinos.
In addition, depending on the final state, 
significant feed-down or leptonic cross-feed backgrounds from misreconstructed excited charm hadron states or $\tau$'s are present. 
Isolation of semitauonic decays from other background processes and the light-lepton final states, 
then requires precise predictions for the kinematics of the signal decay, including all spin correlation and associated interference effects.
The limited size of the available simulated samples
required to account for all these effects
constitutes a dominant uncertainty of the measurements, see e.g.~\cite{Lees:2013uzd,Aaij:2015yra,Huschle:2015rga}.

Whether one is examining the possible effects of NP Wilson coefficients in the context of the LFUV anomalies, 
or one is characterizing the systematic effects inherent to variations of form factor (FF) descriptions of the hadronic matrix elements for precision $|V_{qb}|$ measurement,
(or potentially, both), 
variation of NP Wilson coefficients (NP WCs) and/or FF parameters generically alter decay distributions and acceptances.
Therefore, they modify the signal and possibly background MC templates used in an analysis, 
and thus the inferred values---i.e. preferred regions and best-fit points---of physical parameters 
such as $R(H_c)$ or $|V_{qb}|$.

Consistent interpretation and inference from the data in the context of NP and/or FF variations requires an MC template 
for each considered point in NP WC and FF parameter space.
This approach is sometimes referred to as `forward-folding', and is \emph{prima facie} a computationally prohibitively expensive endeavour. 
Such a program is further complicated because none of the MC generators currently used by the experiments incorporates generic NP  effects, 
nor do they include state-of-the-art treatments of hadronic matrix elements.

The \hammer\ software library provides a solution to these problems: 
A fast and efficient means to reweight large MC samples to any desired NP, or to any description of the hadronic matrix elements.
 \hammer makes use of efficient amplitude-level and tensorial calculation strategies,
and is designed to interface with existing experimental analysis frameworks,
providing detailed control over which NP or hadronic descriptions should be considered. 
The desired reweighting can be implemented either in the event weights or 
in histograms of experimentally reconstructed quantities. 
The only required MC inputs are the event-level ``truth'' four-momenta of existing MC samples. 
Either the event weights and/or histogram predictions may be used, e.g., to generate likelihood functions for experimental fits. 
While \hammer has been designed primarily with $b \to  c \ell \nu$ or $b \to  u \ell \nu$ processes in mind---including not only $B \to \ddss \ell \nu$, 
but also e.g. $\LbLcsln$ or $B_c \to J\!/\!\psi\, \ell \nu$---the general framework of the library has been designed to be extendable 
beyond semileptonic processes \emph{per se}. 

\section{Design overview}

\subsection{Reweighting}

Suppose one has an MC event sample comprising a set of events indexed by $I$, with weights $w_I$ and truth-level kinematics $\{q\}_I$.
Reweighting this sample from an `old' to a `new'  theory requires the truth-level computation of the ratio of the differential rates
\begin{equation}
	\label{eqn:renorm}
	r_I = \frac{ d \Gamma^{\text{new}}_I/d \PSp }{d \Gamma^{\text{old}}_I/d \PSp}\,,
\end{equation}
applied event-by-event via the mapping $w_I \mapsto r_I w_I$. The `old' or `input' or `denominator' theory is typically the SM plus (where relevant) 
a hadronic model --- that is, a form factor (FF) parametrization. (It may also be composed of pure phase space (PS) elements.)
The `new' or `output' or `numerator' theory may involve NP beyond the Standard Model, or a different hadronic model, or both.

Historically, the primary focus of the library is reweighting of $b \to cl\nu$ semileptonic processes,\footnote{%
Throughout, references to $b \to c l \nu$ processes include also $b \to u l \nu$.}
often in multistep cascades such as $B \to D^{(*,**)}(\to DY)\, \tau(\to X\nu) \bar\nu$. 
However, the library's computational structure is designed to allow generalization beyond these particular processes, 
and we therefore frame the following discussion in general terms, before returning to the specific case of semileptonic decays.

\subsection{New Physics generalization}
\label{sec:NPgen}

The \hammer library is designed for the reweighting of processes under theories of the form
\begin{equation}
	\mathcal{L} = \sum_\alpha c_\alpha\, \mathcal{O}_\alpha\,.
\end{equation}
where $\mathcal{O}_\alpha$  are a basis of operators, and $c_\alpha$, are SM or NP Wilson coefficients 
(defined at a fixed physical scale; mixing of the Wilson coefficients under RG evolution, if relevant, must be accounted for externally to the library).
We specify in Table~\ref{tab:NPc}  the conventions used for various $b \to c l \nu$ four-Fermi operators and other processes included in the library.

\enlargethispage{-2.25\baselineskip} 
The corresponding process amplitudes may be expressed as linear combinations $c_\alpha \mathcal{A}_\alpha$. 
They may also be further expressed as a linear sum with respect to a basis of form factors, $F_i$, that encode the physics of hadronic transitions (if any).\footnote{
In all $b \to c$ (or $u$) processes currently handled by \hammer---see Table~\ref{tab:knownampls} for a list---the 
form factors are functions of
\begin{equation}
	q^2 = \big(p_{H_b} -p_{H_c}\big)^2\,,
\end{equation}
or equivalently functions of the recoil,
\begin{equation}
\label{eqn:wdef}
	w = v\cdot v' = \frac{m_{H_b} ^2  + m_{H_c}^2 - q^2}{2m_{H_b}  m_{H_c}}\,,
\end{equation}
with four velocities $v = p_{H_b}  / m_{H_b} $ and $v' = p_{H_c} / m_{H_c}$.
For decays with multi-hadron final states, such as the $\tau \to n \pi \nu$, $n\ge 3$, the form factors are also dependent on multiple invariant masses of the final state hadrons.
Thus $b \to c\tau\nu$ processes with subsequent hadronic $\tau$ decays involve at least two separate sets of hadronic functions at the amplitude level.}
In general, then, an amplitude may be written in the form
\begin{equation}
	\mathcal{M}^{\{s\}}\big(\{q\}\big) = \sum_{\alpha, i} c_\alpha \, F_i\big(\{q\}\big) \, \mathcal{A}^{\{s\}}_{\alpha i}\big(\{q\}\big)\,,
\end{equation}
in which $\{s\}$ are a set of external quantum numbers and $\{q\}$ the set of four-momenta.\footnote{
The momenta of an event passed to the library must all be specified in the same frame. The choice of frame is arbitrary.} 
The object $\mathcal{A}_{\alpha i}$ is an NP- and FF-generalized \emph{amplitude tensor}.  
In the case of cascades, relevant for $B \to D^{(*,**)}(\to DY)\, \tau(\to X\nu) \bar\nu$ decays, 
the amplitude tensor may itself be the product of several subamplitudes, summed over several sets of internal quantum numbers.
The corresponding polarized differential rate
\begin{align}
	\label{eqn:WgtTen}
	\frac{d\Gamma^{\{s\}}}{d\PSp} & = \!\! \sum_{\alpha, i, \beta, j} \!\! c_\alpha c_\beta^\dagger \, F_i F_j^\dagger\!\big(\{q\}\big) \, \mathcal{A}^{\{s\}}_{\alpha i}\mathcal{A}^{\dagger\{s\}}_{\beta j}\!\big(\{q\}\big) \,, \notag \\
	& = \!\! \sum_{\alpha, i, \beta, j} \!\! c_\alpha c_\beta^\dagger \, F_i F_j^\dagger\!\big(\{q\}\big) \,\mathcal{W}^{\{s\}}_{\alpha i \beta j}\,,
\end{align}
in which the phase space differential form $d\PSp$ includes on-shell $\delta$-functions and geometric or combinatoric factors, as appropriate.
\enlargethispage{-2.25\baselineskip} 

The outer product of the amplitude tensor, defined as $\mathcal{W} \equiv \mathcal{A} \mathcal{A}^\dagger$, is a \emph{weight tensor}. 
The object $\sum_{ij} F_i F_j^\dagger \mathcal{W}_{\alpha i \beta j}$ in Eq.~\eqref{eqn:WgtTen} is independent of the Wilson coefficients:
Once this object is computed for a specific $\{q\}$ -- an event -- it can be contracted with any choice of NP to generate an event weight. 
Similarly, on a patch of phase space $\Omega$ --- e.g., the acceptance of a detector or a bin of a histogram --- the marginal rate can now be written as
\begin{equation}
	\Gamma^{\{s\}}_{\Omega} = \sum_{\alpha, \beta} c_\alpha c_\beta^\dagger \int_\Omega d\PSp \, \sum_{ij} F_i F_j^\dagger\big(\{q\}\big)  \mathcal{W}^{\{s\}}_{\alpha i \beta_j} \big(\{q\}\big)\,.
\end{equation}
The Wilson coefficients factor out of the phase space integral, so that the integral itself generates a NP-generalized tensor. 
After it is computed once, it can be contracted with any choice of NP Wilson coefficients, $c_\alpha$, thereafter.

The core of \hammer's computational philosophy is based on the observation that this contraction is computationally much more efficient than the initial computation (and integration).
Hence efficient reweighting is achieved by
\begin{itemize}
	\item Computing NP (and/or FF, see below) generalized objects, and storing them;
	\item Contracting them thereafter for any given NP (and/or FF) choice to quickly generate a desired NP (and/or FF) weight.
\end{itemize}

\subsection{Hadronic generalization}
\label{sec:HG}
Similarly to the NP Wilson coefficients, it is often desirable to be able to generalize variation in the FF parameterization itself. 
For instance, one might contemplate variations along the error eigenbasis of a fit to the FF parameters, or FF parametrizations that are linearized with respect to a basis of parameters, 
such as the BGL FF parametrization~\cite{Grinstein:2017nlq, Boyd:1995sq, Boyd:1997kz} in $B \to D^{(*)} \ell \nu$.
To this end, an FF parametrization with a parameter set $\{\mu\}$ can be linearized around a (best-fit) point, $\{\mu^0\}$ so that 
\begin{equation}
	\label{eqn:FFerr}
	F_{i}\big(\{q\}; \{\mu\}\big) = F_i\big(\{q\}, \{\mu^0\}\big) + \sum_{a} F'_{i,a}\big(\{q\}, \{\mu^0\}\big)\,e_a\,,
\end{equation}
where `$a$' is one or more \emph{variational indices} and $e_{a}$ is the variation. 
In the language of the error eigenbasis case, $F'_{i,a}$ is the perturbation of $F_i$ in the $a$th principal component $e_{a}$ of the parametric fit covariance matrix.

Defining $\xi_{\hat a} \equiv (1, e_a)$ and $\Phi_{i,\hat a} \equiv (F_{i}, F'_{i,a})$, so that Eq.~\eqref{eqn:FFerr} becomes
\begin{equation}
	\sum_{a} \xi_{a} \Phi_{i, a} = F_{i} + \sum_{a'} F'_{i,a'}\,e_{a'}
\end{equation}
then the differential rate,
\begin{equation}
	\label{eqn:GWgtTen}
	\frac{d\Gamma^{\{s\}}}{d\PSp} = \!\! \sum_{\alpha, a, \beta, b} \!\! c_\alpha c_\beta^\dagger \xi_a \xi^\dagger_b \mathcal{U}^{\{s\}}_{\alpha a \beta b}\,,\qquad
	\mathcal{U}^{\{s\}}_{\alpha a \beta b} \equiv \sum_{ij} \Phi_{i,a} \Phi_{j,b}^\dagger\big(\{q\}\big)  \mathcal{W}^{\{s\}}_{\alpha i \beta_j} \big(\{q\}\big)\,,
\end{equation}
with $\mathcal{U}$ an NP- and FF-generalized weight tensor. The $\xi_a$ are independent of $\{q\}$ and factor out of any phase space integral just as the Wilson coefficients do. That is, an integral on any phase space patch,
\begin{equation}
	\Gamma^{\{s\}}_{\Omega} = \sum_{\alpha, \beta, a, b} c_\alpha c_\beta^\dagger \xi_a \xi^\dagger_b \int_\Omega d\PSp \,\, \mathcal{U}^{\{s\}}_{\alpha a \beta b}\,. 
\end{equation}
One may thus tensorialize the amplitude with respect to Wilson coefficients and/or FF linearized variations, to be contracted later with with NP or FF variation choices 
(the latter within the regime of validity of the FF linearization).
Hereafter, the $\xi_a$ are referred to as `FF uncertainties' or `FF eigenvectors' following the nominal fit covariance matrix example. 

\subsection{Rates}

In certain use cases, it is also useful to compute and fold in an overall ratio of rates $\Gamma^{\text{old}}/\Gamma^{\text{new}}$, 
or the rates themselves, $\Gamma^{\text{new}, \text{old}}$, may be required. 
For example, if the MC sample has been initially generated with a fixed overall branching ratio, $\BR_{\text{new}}$, 
one might wish to enforce this constraint via an additional multiplicative factor $\BR_{\text{old}}/\BR_{\text{new}}$.

The different components computed by \hammer are then: 
\begin{enumerate}
\item The NP- and/or FF-generalized tensor for $(d \Gamma^{\text{new}}_I/d \PSp) / (d \Gamma^{\text{old}}_I/d \PSp)$, 
via Eq.~\eqref{eqn:GWgtTen}, noting that the denominator carries no free NP or FF variational index. (The ratio $r_I$ is then itself generally at least a rank-2 tensor.); 
\item The NP- and/or FF-generalized \emph{rate tensors} $\Gamma^{\text{old, new}}$, which need be computed only once for an entire sample.  
(These rates require integration over the phase space, which is achieved by a dedicated multidimensional Gaussian quadrature integrator.)
\end{enumerate}

\subsection{Primary code functionalities}

The computational core of \hammer calculates the NP or FF generalized tensors event-by-event for any process known to the library (see Tab.~\ref{tab:knownampls} for a list), 
and as specified by initialization choices (more detail is provided in Sec.~\ref{sec:code_flow}) and specified form factor parametrizations.
This core is supplemented by a wide array of functionalities to permit manipulation the resulting NP- and FF-generalized weight tensors as needed. 
This may include binning --- equivalent to integrating on a phase space patch --- the weight tensors into a histogram of any desired reconstructed observables, 
and/or it may include folding of detector simulation smearings, etc.
Such histograms have NP- and FF-generalized tensors as bin entries, and we therefore call them \emph{generalized} or \emph{tensor} histograms.
Once such NP- and FF-generalized tensor objects are computed and stored, contraction with NP or FF eigenvector choices 
permits the library to efficiently generate actual event weights or histogram bin weights for any theory of interest.

The architecture of \hammer is designed around several primary functionalities:
\begin{enumerate}
	\item Provide an interface to determine which processes are to be reweighed, and which (possibly multiple) schemes for form factor parametrizations are to be used. 
	This includes handling for (sub)processes that were generated as pure phase space.
	\item Parse events into cascades of amplitudes known to the library, and compute their corresponding NP- and/or FF-generalized amplitude or weight tensor, 
	as well as the respective rate tensors, as needed.
	\item Provide an interface to generate histograms (of arbitrary dimension), and bin the event weight tensors --- i.e., $r_I w_I$, as in Eq.~\eqref{eqn:renorm} --- 
	into these histograms, as instructed. 
	This includes functionality for weight-squared statistical errors, functionality for generation of \texttt{ROOT} histograms, 
	as well as extensive internal architecture for efficient memory usage.
	\item Efficiently contract generalized weight tensors or bin entries against specific FF variational or NP choices, to generate an event or bin weight. 
	This includes extensive internal architecture to balance speed versus memory requirements.
	\item Provide interface to save and reload amplitude or weight tensors or generalized histograms, 
	to permit quick reprocessing into weights from precomputed or `initialized' tensor objects. 
\end{enumerate}
Examples of the implementation of these functionalities is shown in extensive examples provided with the source code.

\subsection{Code flow}
\label{sec:code_flow}

A \hammer program may have two different types of structure: An \emph{initialization} program, so called as it runs on MC as input, and may generate \hammer format files; 
or a \emph{analysis} program, which may reprocess histograms or event weights that have already been saved in an initialization run. 

An \emph{initialization} program has the generic flow:
\begin{enumerate}
	\item Create a \textlst{Hammer} object.
	\item Declare included or forbidden processes, via \textlst{includeDecay} and \textlst{forbidDecay}.
	\item Declare form factor schemes, via \textlst{addFFScheme} and \textlst{setFFInputScheme}.
	\item (Optional) Add histograms, via \textlst{addHistogram}.
	\item (Optional) Declare the MC units, via \textlst{setUnits}. 
	\item (Optional) Declare partially-specialized WC subspaces, via \textlst{createWCSpecialization} etc.
	\item Initialize the \textlst{Hammer} class members with \textlst{initRun}.
	\item (Optional) Change FF default settings with \textlst{setOptions}, or (if not SM) declare the Wilson coefficients for the input MC via \textlst{setWilsonCoefficients}.
	\item (Optional) Fix FF uncertainty choice to special choices in weight calculations (histogram binnings), via \textlst{specializeFFInHistogram}.
	\item Each event may contain multiple processes, e.g., a signal and tag $B$ decay. Looping over the events:
		\begin{enumerate}
			\item Initialize event with \textlst{initEvent}. For each process in the event:
			\begin{enumerate}
				\item Create a \textlst{Process} object.
				\item Add particles and decay vertices to create a process tree, via \textlst{addParticle} and \textlst{addVertex}.
				\item Decide whether to include or exclude processes from an event via \textlst{addProcess} and/or \textlst{removeProcess}.
			\end{enumerate}
			\item Compute or obtain event observables -- specific particles can be extracted with \textlst{getParticlesByVertex} or other programmatic means -- 
			and specify the corresponding histogram bins to be filled via \textlst{fillEventHistogram}.
			\item Initialize and compute the process amplitudes and weight tensors for included processes in the event, and fill histograms with event tensor weights -- 
			the direct product of include process tensor weights -- via \textlst{processEvent}.
			\item  (Optional) Save the weight tensors for each event, with \textlst{saveEventWeights} to a buffer.
			\item  (Optional) Save the rate tensors, with \textlst{saveRates} to a buffer.
		\end{enumerate}
	\item (Optional) Produce histograms with \textlst{getHistogram(s)} and/or save them to a buffer with \textlst{saveHistograms}. 
	NP choices are implemented with \textlst{setWilsonCoefficients}, FF variations are set with \textlst{setFFEigenvectors}.
	\item (Optional) Save an autogenerated \texttt{bibTeX} list of references used in the run with \textlst{saveReferences}.
\end{enumerate}

By contrast, an \emph{analysis} program (from a previously initialized sample, stored in a buffer) has the generic flow:
\begin{enumerate}
	\item Create a \textlst{Hammer} object and specify the input file.
	\item Load or merge the run header --- include or forbid specifications, FF schemes, or histograms --- with \textlst{loadRunHeader} 
	(before \textlst{initRun}). One may further declare additional histograms to be compiled (from saved event weight data) via \textlst{addHistogram}
	(\textlst{loadRunHeader} must be called before \textlst{initRun} to ensure newly-added histograms can access previously saved form factor schemes).
	\item (Optional) Load or merge saved histograms with \textlst{loadHistograms}, and/or generate desired histograms with \textlst{getHistogram(s)}. NP choices are implemented with \textlst{setWilsonCoefficients}.
	\item (Optional) Looping over the events:
		\begin{enumerate}
			\item Initialize event with \textlst{initEvent}.
			\item If desired, remove processes from an event with \textlst{removeProcess}.
			\item Reload event weights with \textlst{loadEventWeights}.
			\item Specify histograms to be filled via \textlst{fillEventHistogram}.
			\item Fill histograms with event weights via \textlst{processEvent}.
		\end{enumerate}
\end{enumerate}
A summary of placement relative to \textlst{initRun} for a selection of configuration methods is provided in Table~\ref{tab:initrun}.

\begin{table}[h]
\newcolumntype{M}{ >{\CPPidentifierstyle}m{7.5cm}<{}}
\def\simplecollect#1#2\ignorespaces#3\unskip{#1{#3}\unskip}
\newcommand*{\textlsttab}[1]{\raggedright{\textlst{#1}}}
\newcolumntype{S}{>{\simplecollect\textlsttab} m{7.5cm}}
\newcolumntype{C}{ >{\centering\arraybackslash} m{2cm} <{}}
\renewcommand{\arraystretch}{0.85}
\begin{tabular}{SCC}
\hline
\multicolumn{1}{l}{\multirow{2}{*}{Configuration method}} & \multicolumn{2}{c}{Place versus \textlst{initRun} } \\
\multicolumn{1}{c}{}	& \textbf{before} & \textbf{after}\\
\hline\hline
includeDecay & \cmark & \\
forbidDecay & \cmark & \\
addPurePSVertices & \cmark & \\
\hline
addFFScheme & \cmark & \\
setFFInputScheme & \cmark & \\
renameFFEigenvectors & \cmark & \cmark \\ 
\hline
addHistogram & \cmark & \\
keepErrorsInHistogram & \cmark & \\
addTotalSumOfWeights & \cmark & \\
collapseProcessesInHistogram  & \cmark & \\
\hline
createWCSpecialization & \cmark & \\
setWCSpecializationOrigin|Basis|Coord & \cmark& \\
applyWCSpecializationInWeights|Histogram(s)| & \multirow{2}{*}{\cmark} & \multirow{2}{*}{\tikz[baseline={([yshift=-0.6ex]current bounding box.center)}]{\node (app) [inner sep = 0pt] {\cmark\rlap{\footnotemark[1]}}}} \\
 !\qquad\lstinline!AllHistograms &&\\
specializeFFInHistogram & & \cmark \\
\hline
setUnits & \cmark & \\
setOptions & \cmark\rlap{\footnotemark[2]} & \cmark \\
loadRunHeader & \cmark & \\
\hline
\end{tabular}
\footnotetext[1]{\fontsize{8}{8}\selectfont Requires \lstinline[basicstyle=\ttfamily]{reconcileSpecializations}.}
\footnotetext[2]{\fontsize{8}{8}\selectfont Settings for user-created objects, such as duplicated FF classes, must be called after their creation.\\\noindent Certain default options are set by \lstinline[basicstyle=\ttfamily]{initRun} itself, which may be adjusted afterwards, e.g. specialization logic. See main text.}
\caption{Placement relative to \textlst{initRun()} initialization for a selection of methods used to configure a \hammer program in nominal \hammer code flows.}
\label{tab:initrun} 
\end{table}

\section{The Hammer forge}
In the following, we describe various core parts of the \hammer Application Programming Interface (API). This includes a detailed explanation of information handling and rules enforced by the computational core in following user specifications, assembling amplitudes, or returning histograms or weights, among other functionalities.
The library itself is implemented in \texttt{C++}, along with a \texttt{Python3} wrapper of the API with \texttt{cppyy}, that uses identical syntax. If \hammer is installed with ROOT support, the ROOT internal \texttt{cppyy} will be automatically used, allowing smooth interoperability with \texttt{pyROOT}. A legacy \texttt{Python3} wrapper using \texttt{Cython}, with slightly different syntax (``snake case'' instead ``camel case'') and no \texttt{pyROOT} support\footnote{For example, direct tensor histogram evaluation to \textlst{TH\{1,2,3\}D} objects is not available with the \texttt{Cython} bindings.}, is still present, albeit deprecated and may be removed in future \hammer versions.

To facilitate the integration with existing statistical analysis packages such as \texttt{RooFit} and \texttt{pyhf}, the community have developed external wrappers around \hammer, 
such as \texttt{RooHammerModel}~\cite{GarciaPardinas:2020yrd} and \texttt{Redist-HAMMER}~\cite{colonna2025}. 
We refer the interested reader to their documentation for information on their usage, internal architecture and which \hammer functionalities they encapsulate.

\begin{figure}[t]
	\includegraphics[width = \textwidth]{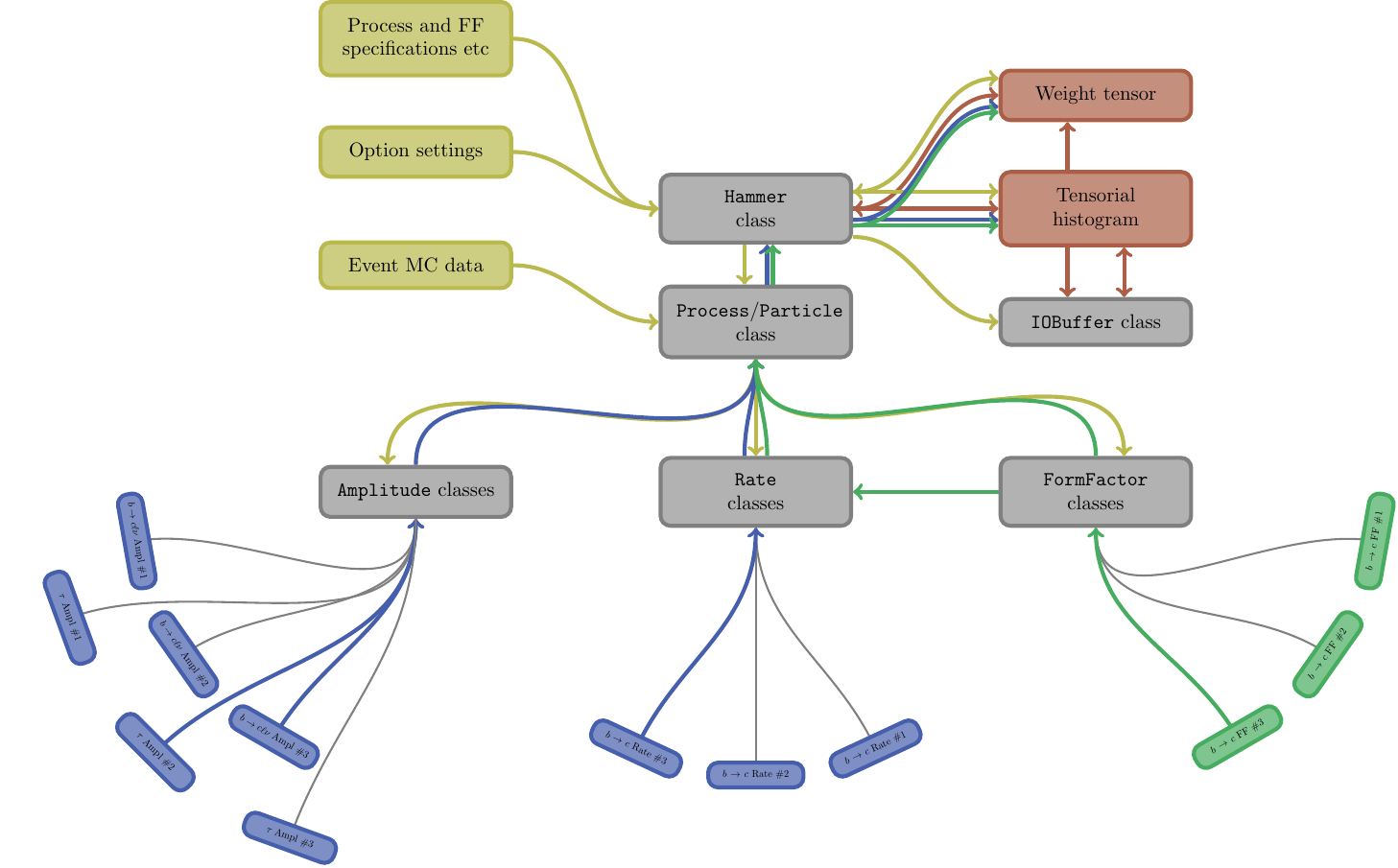}
	\caption{Schematic architecture of \hammer. The flow of user-specified choices or event data is shown by yellow arrows. Blue (green) arrows denote the flow of computed
	information, in particular amplitude, weight or rate (form factor) tensors. 
	Red arrows highlight the flow of \hammer output, which may be saved or reloaded. Most internal \hammer classes are not shown in this schematic.}
	\label{fig:Scheme}
\end{figure}

We will consider here the \texttt{C++} interface only; the discussion is ordered by scope, rather than the typical code flow. 
The library provides five main core classes in its user-facing interface: the \textlst{Hammer} class itself; the \textlst{Process}, \textlst{Particle}, and \textlst{FourMomentum} classes, used to create events; 
and the \textlst{IOBuffer} class (with its list container \textlst{IOBuffers}) used for saving and loading precomputed objects. 
Internal computational classes include \textlst{Amplitude}, \textlst{FormFactor} 
and \textlst{Rate} classes, that encode the physics of processes known to the library. A schematic of the architecture of \hammer is shown in Fig.~\ref{fig:Scheme}.

\subsection{From the process tree to an amplitude tensor}
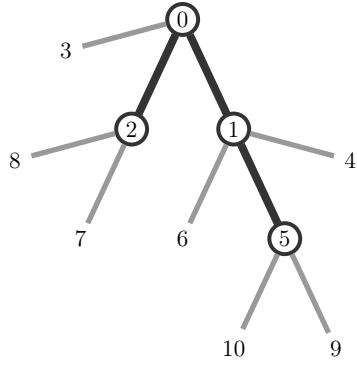
\begin{figure}[t]
\begin{tikzpicture}
		\tikzset{every node/.style={draw=black!80,line width=1.5pt,fill=white,circle,inner sep=2pt,minimum size = 12pt,align=center, scale = 0.8, transform shape}}
		\tikzstyle{level 1}=[sibling angle=50, level distance = 0.8*2cm];
		\tikzstyle{level 2}=[sibling angle=50, level distance = 0.8*2cm];
		\tikzstyle{level 3}=[sibling angle=50, level distance = 0.8*2cm];
		\tikzstyle{nil}=[draw=none]
		\tikzstyle{edge from parent}=[draw=black!40, line width = 2pt]
		\node (P) {0} [grow cyclic,shape=circle,clockwise from=-65]
			child  { [grow cyclic,shape=circle,clockwise from=-15] node {1} edge from parent [draw = black!80, line width = 3pt]
				child  {node [nil] {4}}
				child  {[grow cyclic,shape=circle,clockwise from=-65]  node {5} edge from parent [draw = black!80, line width = 3pt]
					child {node [nil] {9}}
					child {node [nil] {10}}
				}
				child  {node [nil] {6}};
			}
			child  {[grow cyclic,shape=circle,clockwise from=-115] node {2} edge from parent [draw = black!80, line width = 3pt]
				child  {node [nil] {7}}
				child  {node [nil] {8}};
			}
			child  {node [nil] {3}};
\end{tikzpicture}
\caption{Example process tree for a decay cascade involving 10 particles (numbers), 4 vertices (circles) and 3 edges (dark lines).}
\label{fig:processtree}
\end{figure}

A typical decay cascade is contained in the library by the \textlst{Process} class; an event may contain multiple \textlst{Process} instances as e.g., is the case for a signal plus tag $B$-$\bar{B}$ pair. 
Each cascade may be simply represented in graphical terms as a `process tree', as shown in Fig.~\ref{fig:processtree}: Each decay vertex is labelled by its local parent particle,
connected to subsequent daughter decays by  an edge (i.e. a line, or formally, a propagator).
Each particle in the cascade is itself assigned an index, and then the decay vertex is represented as a map from a parent index, to the indices of all its daughters. 

\hammer assembles the process tree through two methods \textlst{Process::addParticle} and \textlst{Process::addVertex}. The former adds a \textlst{Particle} class object --  a momentum and a PDG code -- to a container of particles; the latter fills a map of each parent to its daughters for each decay vertex. In the case of Fig.~\ref{fig:processtree}, the first two vertices of the cascade may be built explicitly as follows:
\begin{activett}
	Process proc;
	size_t idx0 = proc.addParticle(Particle{{E_0, px_0, py_0, pz_0}, pdg_0});
	size_t idx1 = proc.addParticle(Particle{{E_1, px_1, py_1, pz_1}, pdg_1});
	size_t idx2 = proc.addParticle(Particle{{E_2, px_2, py_2, pz_2}, pdg_2});
	size_t idx3 = proc.addParticle(Particle{{E_3, px_3, py_3, pz_3}, pdg_3});
	size_t idx7 = proc.addParticle(Particle{{E_7, px_7, py_7, pz_7}, pdg_7});
	size_t idx8 = proc.addParticle(Particle{{E_8, px_8, py_8, pz_8}, pdg_8});
	
	proc.addVertex(idx0, {idx1,idx2,idx3});
	proc.addVertex(idx2, {idx7,idx8});
\end{activett}
and so on. Particles and vertices need not be added in order; helper functions are provided in the code examples for automatically parsing HepMC files.

From the filled process tree, \hammer determines several hashes or sets of hashes, that encode the structure of the tree: In particular, i) a set of the hashes of parent and daughter particle PDG codes at each vertex; ii) a combined hash for the process -- a `process ID' -- providing a $1$-$1$ identifier between the full decay cascade and a \textlst{size\_t} integer. For any process, the latter can be obtained by the method \textlst{Process::getId}. The former will be relevant later for understanding how `included' and `forbidden' processes are identified.

At this stage, the natural computational step is to map each vertex into a corresponding amplitude tensor, contracting exchanged quantum numbers along each edge to form a single tensor for the whole process tree. In the simplest cases, this is precisely the strategy adopted by \hammer, i.e., the particle ID hashes constructed at each vertex are looked up in a dictionary of the signatures of available \textlst{Amplitude} classes. A similar technique, using the hash of the hadronic particles in a vertex, is used to identify whether form factors are needed at each vertex. (If form factors are required at a vertex, \hammer will obtain the relevant form factor parameterization as specified by the user for the hadronic transition in question.) If no amplitude is found for a vertex, \hammer will simply skip this step of the cascade. This behavior means that hammer implicitly prunes potentially highly extended cascades, providing an amplitude tensor only for vertices \hammer `knows' (i.e., the parts of the cascade we care about for understanding NP effects or FF parametrizations). 

In certain cases, the strategy adopted for determining the process amplitude is more sophisticated than a vertex-by-vertex approach. For certain decays, it can be computationally advantageous to calculate an amplitude for two adjacent amplitudes. For example, in $B \to (D^* \to D\gamma) \ell \nu$, simpler expressions can be obtained if one calculates the entire `merged' amplitude, treating the $D^*$ as an on-shell internal state, rather than two separate amplitudes exchanging $D^*$ spin. Similarly, for $\tau \to (\rho  \to \pi\pi) \nu$, treatment of non-resonant effects from the broad $\rho$ motivates expressing this amplitude as one merged amplitude, even though in the process tree it would be represented as two vertices. Multistep decays involving the broad $D^{**}$ may also be more tractable when merged in this manner. Thus, in addition to vertex amplitudes, \hammer is also capable of processing `edge' amplitudes, that is, one amplitude belonging to two adjacent vertices connected by an edge in the process tree. It can therefore happen that although \hammer does not know the amplitude for a particular vertex, it does know an edge amplitude involving that vertex and another.

To explain what this means in practice for the user, it's useful to introduce a vertex and edge notation for the process tree. If \hammer knows the amplitude at a vertex, the vertex is denoted by a filled circle, and if unknown, by an open circle. If an edge vertex is available for two vertices, we connect them by a double line.  This leads to five different types of amplitude combinations, defined in Table~\ref{tab:ampltypes}. The arithmetic followed by \hammer in determining the amplitude from the tree is as follows:
\begin{enumerate}
	\item Fill all available pure edges by lowest (i.e., furthest from head parent) to highest depth in the process tree, being sure not to assign the same vertex twice
	\item Repeat for partial then full edges
	\item Assign known vertex amplitudes to any remaining free vertices.
\end{enumerate}
Two examples or this arithmetic are shown in Table~\ref{tab:examplearith}.

\begin{table}[h]
\newcolumntype{C}{ >{\centering\arraybackslash} m{2cm} <{}}
\renewcommand{\arraystretch}{1.2}
\begin{tabular}{c|CC|CCC}
	\hline
	& \multicolumn{2}{c|}{Vertex}  & \multicolumn{3}{c}{Edge}  \\
	\hline\hline
	Amplitude & Known & Unknown & Pure & Partial & Full \\
	Notation & \cvertex & \overtex & \puredge & \paredge & \fuledge \\
	\hline
\end{tabular}
\caption{Definition of vertex and edge amplitude types.}
\label{tab:ampltypes}
\end{table}

\begin{table}[t]
\newcolumntype{C}{ >{\centering\arraybackslash} m{5cm} <{}}
\begin{tabular}{C|C}
\hline
Known Amplitudes & Evaluated Amplitudes \\
\hline\hline
\vspace{5pt}
\begin{tikzpicture}
		\tikzset{every node/.style={scale = 0.7, transform shape}}
		\tikzstyle{level 1}=[sibling angle=50, level distance = 0.7*2cm];
		\tikzstyle{level 2}=[sibling angle=50, level distance = 0.7*2cm];
		\tikzstyle{level 3}=[sibling angle=50, level distance = 0.7*2cm];
		\tikzstyle{o}=[draw=black!80,line width=1.5pt,fill=white,circle,inner sep=2pt,minimum size = 12pt,align=center, text=black]
		\tikzstyle{c}=[draw=black!80,line width=1.5pt,fill=black!80,circle,inner sep=2pt,minimum size = 12pt,align=center, text=white]
		\tikzstyle{edge from parent}=[draw=black!40, line width = 2pt]
		\node (P) [c] {0} [grow cyclic,shape=circle,clockwise from=-65]
			child  { [grow cyclic,shape=circle, clockwise from=-65] node [o] {1} edge from parent [draw = black!80, line width = 1.5pt, double]
				child  {node [c] {5} edge from parent [draw = black!80, line width = 2pt]
				}
			}
			child  {node [c] {2} edge from parent [draw = black!80, line width = 2pt]
			};
\end{tikzpicture}
& \vedge{0}{1}, \vvertex{2}, \vvertex{5} \\
\hline
\vspace{5pt}
\begin{tikzpicture}
		\tikzset{every node/.style={scale = 0.7, transform shape}}
		\tikzstyle{level 1}=[sibling angle=50, level distance = 0.7*2cm];
		\tikzstyle{level 2}=[sibling angle=50, level distance = 0.7*2cm];
		\tikzstyle{level 3}=[sibling angle=50, level distance = 0.7*2cm];
		\tikzstyle{o}=[draw=black!80,line width=1.5pt,fill=white,circle,inner sep=2pt,minimum size = 12pt,align=center, text=black]
		\tikzstyle{c}=[draw=black!80,line width=1.5pt,fill=black!80,circle,inner sep=2pt,minimum size = 12pt,align=center, text=white]
		\tikzstyle{edge from parent}=[draw=black!40, line width = 2pt]
		\node (P) [c] {0} [grow cyclic,shape=circle,clockwise from=-65]
			child  { [grow cyclic,shape=circle,clockwise from=-65] node [o] {1} edge from parent [draw = black!80, line width = 1.5pt, double]
				child  {node [o] {5} edge from parent [draw = black!80, line width = 1.5pt, double]
				}
			}
			child  {node [c] {2} edge from parent [draw = black!80, line width = 2pt]
			};
\end{tikzpicture}
&  \vvertex{0}, \vedge{1}{5}, \vvertex{2} \\
\hline
\end{tabular}
\caption{Example arithmetic for filling amplitudes for the process tree of Fig.~\ref{fig:processtree}, assuming different example sets of known amplitudes in \hammer.}
\label{tab:examplearith}
\end{table}

\subsection{Available amplitudes and form factor parametrizations}
\label{sec:Proc}
The list of vertex and edge amplitudes known to \hammer is shown in Table~\ref{tab:knownampls}. Also shown are correspondingly available form factor parametrizations, as appropriate.
A full list of the settable form factor parameters and switches, and their default values, can be retrieved programmatically in YAML format via the method \textlst{Hammer::saveOptionCard()}, accepting the file name and a boolean option of whether to save the current or default values of the settings.

The full list of form factor parametrizations known to the library can be displayed by \textlst{Hammer::showAvailableFFParams()}, 
organized by form factor class prefix string (defined in Sec.~\ref{sec:FFset} below). 
This method may also take a class prefix string (such as \textlst{"BtoD*"}) as an argument, 
displaying the known form factors known for the processes associated with that prefix.

\begin{table}[tb]
\begin{center}
\renewcommand{\arraystretch}{0.95}
\newcolumntype{D}{ >{\centering\arraybackslash} m{0.31\linewidth} <{}}
\newcolumntype{E}{ >{\raggedright\arraybackslash\quad\ttfamily} m{0.55\linewidth} <{}}
\newcommand{\citerm}[1]{{\rmfamily{\cite{#1}}}}
\scalebox{0.95}{\parbox{\linewidth}{
\begin{tabular}{D|E}
\hline\hline
Process &  \textrm{FF parametrizations}\\
\hline
$B \to D^{(*)} \ell \nu$  
	&  ISGW2$^*$\,\citerm{Scora:1995ty,Isgur:1988gb}, BGL$^{*\ddagger}$\,\citerm{Grinstein:2017nlq, Boyd:1995sq, Boyd:1997kz}, CLN$^{*\ddagger}$\,\citerm{Caprini:1997mu}, \\
	& BLPR$^{\ddagger}$\,\citerm{Bernlochner:2017jka}, BLPRXP$^{\ddagger}$\,\citerm{Bernlochner:2022ywh}\\
$B \to (D^* \to D \pi) \ell \nu$  & ISGW2$^*$, BGL$^{*\ddagger}$, CLN$^{*\ddagger}$, BLPR$^{\ddagger}$, BLPRXP$^{\ddagger}$\\
$B \to (D^* \to D\gamma) \ell \nu$  &  ISGW2$^*$, BGL$^{*\ddagger}$, CLN$^{*\ddagger}$, BLPR$^{\ddagger}$, BLPRXP$^{\ddagger}$\\
$ B \to D^*_0 \ell \nu$ &  ISGW2$^*$, LLSW$^*$\,\citerm{Leibovich:1997em, Leibovich:1997tu}, BLR$^{\ddagger}$\,\citerm{Bernlochner:2017jxt, Bernlochner:2016bci}\\
$ B \to D^*_1 \ell \nu$ &  ISGW2$^*$, LLSW$^*$, BLR$^{\ddagger}$ \\
$ B \to D_1 \ell \nu$ &  ISGW2$^*$, LLSW$^*$, BLR$^{\ddagger}$ \\
$ B \to D^*_2 \ell \nu$ &  ISGW2$^*$, LLSW$^*$, BLR$^{\ddagger}$ \\
$ B \to (\rho \to \pi\pi)\ell \nu$ & ISGW2$^*$, BSZ$^{\ddagger}$\,\citerm{Straub:2015ica}\\
$ B \to (\omega \to \pi\pi\pi)\ell \nu$ &  ISGW2$^*$, BSZ$^{\ddagger}$\\
$\Lambda_b \to \Lambda_c \ell \nu$ & PCR$^*$\,\citerm{Pervin:2005ve}, BLRS$^{\ddagger}$\,\citerm{Bernlochner:2018bfn, Bernlochner:2018kxh}, BLRSXP\,\citerm{Bernlochner:2023jkp}$^{\ddagger}$ \\
$\Lambda_b \to \Lambda_c^* \ell \nu$ & PCR$^*$, LSPR$^{\ddagger}$\citerm{Leibovich:1997az,Papucci:2021pmj}\\
$ B_c \to (J\!/\!\psi \to \ell\ell)\ell \nu$ & Kiselev$^*$\,\citerm{Kiselev:2002vz}, EFG$^*$\,\citerm{Ebert:2003cn}, BGL$^{*\ddagger}$\,\citerm{Cohen:2019zev}, $\ldots$ \\
$B \to \pi \ell \nu$ & ISGW2$^*$, BCL$^{*\ddagger}$\,\citerm{Lattice:2015tia}, GKvD\,\citerm{Gubernari:2018wyi}\\
$B_s \to K \ell \nu$ & ISGW2$^*$, BCL$^{*\ddagger}$\citerm{FlavourLatticeAveragingGroupFLAG:2021npn} \\
$B_s \to D_s \ell \nu$ & BCL$^{*\ddagger}$\citerm{McLean:2019qcx}\\
\hline
$\tau \to \pi \nu$ & \textrm{---}\\
$\tau \to \ell \nu \nu$ &   \textrm{---}\\
$\tau \to 3\pi \nu$ &  RCT$^*$\,\citerm{Kuhn:1992nz,Shekhovtsova:2012ra,Nugent:2013hxa}\\
\hline
$D_1 \to (D^* \to D \pi/\gamma)\pi$ & PW \\
$D_2^* \to (D^* \to D \pi/\gamma)\pi$ & PW \\
$D_2^* \to D\pi$ & PW \\
\hline
\multicolumn{2}{c}{Planned for future release}\\
\hline
$ B_{(c)} \to \ell \nu$ &  MSbar\\
\hline
$\tau \to 4\pi \nu$ &  RCT$^*$ \\
$\tau \to (\rho \to \pi\pi)\nu$ &  \textrm{---}\\
\hline\hline
\end{tabular}
}}
\caption{Implemented amplitudes in \hammer and corresponding form factor parametrizations. 
SM-only parametrizations are indicated by a $*$ superscript. Form factor parameterizations that include linearized variations are denoted with a $\ddagger$ superscript.  
These are named in the library by adding a ``\textlst{Var}'' suffix, e.g. ``\textlst{BGLVar}''. 
\textbf{For each $b \to c$ process, also included are analogous $bs \to cs$ processes with the same form factor parameterizations.}
Similarly, charmed meson cascades to $D$ and $\pi$'s include charm-strange equivalent processes to final states containing $D_s$ or $K$.
The method \textlst{showAvailableFFParams} displays the latest list of FF parametrizations known to the library.}
\label{tab:knownampls}
\end{center}
\end{table}

\subsection{Including and excluding processes}
\label{sec:allowforbid}
The \hammer library contains an interpreter between a string representation of a vertex and the corresponding PDG codes of incoming and outgoing particles. At present, a string representation of a vertex, or `vertex string', is formed  by concatenating a single parent name with daughter names, in the form \textlst{ParentDaughter1Daughter2...}. The interpreter uses the syntax that particle names are parsed by a capital letter: the full list of names is provided in Table~\ref{tab:particlespecs}. The interpreter maps a vertex string to all possible \emph{charge conserving} processes allowed by the charges of the specified particle names. For example the vertex string \textlst{"D*DPi"} is interpreted as all twelve possible $D^* \to D\pi$ vertices, while \textlst{"D*+DPi"} is interpreted as only the $D^{*+} \to D^+\pi^0$, $D^{*+} \to D^0\pi^+$, and (the heavily CKM suppressed) $D^{*+} \to \bar{D}^0\pi^+$ decays, and finally the vertex string \textlst{"D*+D0Pi"} corresponds to the unique decay $D^{*+} \to D^0\pi^+$.

The decay processes to be reweighed by \hammer are specified via \textlst{Hammer::includeDecay}, which takes a vector of vertex strings $\{V_1,V_2,\ldots, V_n\}$ as an argument, and may be invoked multiple times.
Each \textlst{includeDecay} specification is \emph{inclusive} and permits any process tree whose set of vertices $P$ \emph{contains} $\{V_1,V_2,\ldots, V_n\}$. The boolean logic applied by \textlst{includeDecay} is \textlst{AND} between each vertex string element, and \textlst{OR} between separate invocations of \textlst{includeDecay}. For example
\begin{activett}
	ham.includeDecay({"BD*TauNu", "D*DGamma"});
	ham.includeDecay({"BDMuNu"});
\end{activett}
means `Reweigh a process that either contains vertices ($B \to D^*\tau\nu$ \textbf{and} $D^* \to D\gamma$) \textbf{or} the vertex ($B \to D \mu \nu$)'. Hence e.g. $\bar{B}^0 \to (D^{*+} \to (D^+ \to K^+ \pi^+ \pi^-)\gamma)(\tau^- \to \ell^-\nu\nu)$ would be included. Radiative photons are automatically accounted for, and need not be specified in \textlst{includeDecay} specifications (see Sec. \ref{sec:photos}).

Processes are forbidden with the \textlst{Hammer::forbidDecay} method, which similarly takes a vector of vertex strings $\{V_1,V_2,\ldots, V_n\}$, and employs the same boolean structure as \textlst{includeDecay}. However, \textlst{forbidDecay} specifications are \emph{exclusive} and forbids only process trees whose set of vertices $P$ \emph{equals} $\{V_1,V_2,\ldots, V_n\}$. For example
\begin{activett}
	ham.forbidDecay({"B+D0barMuNu"});
\end{activett}
means `Exclude a process that contains only the vertex $B^+ \to \bar{D}^0 \mu^+ \nu_\mu$', but e.g., this would not exclude a process involving a subsequent $D$ decay.

Inclusion or exclusion of processes may also be specified via an initialization card in YAML format. For example, the equivalent to the above \textlst{includeDecay} and \textlst{forbidDecay} invocations is
\begin{activett}
Include: [ [ BD*TauNu, D*Dpi ], BDMuNu ]
Forbid: [ B+D0barMuNu ]
\end{activett}
using the same vertex string syntax and symbology.

\begin{table}[t]
\renewcommand{\arraystretch}{0.9}
\newcolumntype{C}{ >{\centering\arraybackslash\ttfamily} m{2cm} <{}}
\newcolumntype{D}{ >{\centering\arraybackslash $} m{5.5cm} <{$}}
\scalebox{0.7}{\parbox{\linewidth}{
\begin{tabular}[t]{|C|D|}
	\hline
	\textrm{Symbol} & \text{Particle(s)} \\
	\hline\hline
        D       & D^+\,,\quad D^-\,,\quad D^0\,,\quad \bar{D}^0 \\
        D*   & D^{*0}\,,\quad D^{*-}\,,\quad D^{*+}\,,\quad \bar{D}^{*0} \\
        Lc &  \Lambda^+_c\,, \Lambda^-_c \\
        B       & B^0\,,\quad B^-\,,\quad B^+\,,\quad \bar{B}^0 \\
        Bs       & B_s^0\,,\quad \bar{B}_s^0 \\
        Lb 	&  \Lambda^0_b\,, \bar{\Lambda}^0_b \\
        Bc 	& B_c^-\,,\quad B_c^+\\
        K       & K^+\,,~K^-\,,~K_L^0\,,~K_S^0\,,~K^0\,,~\bar{K}^0  \\
        Pi      & \pi^0\,,\quad \pi^+\,,\quad \pi^- \\
        D**0* &  D_0^{*0}\,, \quad D_0^{*-}\,, \quad D_0^{*+}\,, \quad \bar{D}_0^{*0} \\
        D**1* & D_1^*\,, \quad D_1^{*-}\,, \quad D_1^{*+}\,, \quad \bar{D}_1^{*0} \\
        D**1 & D_1^0\,, \quad D_1^{-}\,, \quad D_1^{+}\,, \quad \bar{D}_1^0 \\
        D**2* & D_2^{*0} \quad D_2^{*-}\,, \quad D_0^{*+}\,, \quad \bar{D}_0^{*0} \\
        Lc*2595 &  \Lambda^{*+}_c(2595)\,, \Lambda^{*-}_c(2595) \\
        Lc*2625 &  \Lambda^{*+}_c(2625)\,, \Lambda^{*-}_c(2625) \\
        Jpsi & J/\psi\\
        E       & e^+\,,\quad e^- \\
        Mu      & \mu^-\,,\quad \mu^+ \\
        Tau     & \tau^-\,,\quad \tau^+ \\
	Nu      & \nu_e\,,\quad \bar\nu_e\,,\quad \nu_\mu\,,\quad \bar\nu_\mu\,,\quad \nu_\tau\,,\quad \bar\nu_\tau \\
        Ell     & \mu^-\,,\quad \mu^+\,,\quad e^-\,,\quad e^+ \\
        W       & W^+\,,\quad W^- \\   
        \hline
        Gamma   & \gamma \\
        Tau+    & \tau^+ \\
        E+      & e^+ \\
        Mu+     & \mu^+ \\
        K+      & K^+ \\
        K0S     & K_S^0 \\
        K0L     & K_L^0 \\
        K0     & K^0 \\
        K0bar     & \bar{K}^0 \\
        B0      & B^0 \\
        B0bar   & \bar{B}^0 \\
        B+      & B^+ \\
        B+-     & B^+\,,\quad B^- \\
        Babar   & B^0\,,\quad \bar{B}^0 \\
        Bs0      & B_s^0 \\
        Bs0bar   & \bar{B}_s^0 \\   
        Bc+      & B_c^+ \\
        D0      & D^0 \\
        D+      & D^+ \\
        D0bar   & \bar{D}^0 \\   
        \hline
\end{tabular}
\begin{tabular}[t]{|C|D|}
	\hline
	\textrm{Symbol} & \text{Particle(s)} \\
	\hline\hline
	D+-     & D^+\,,\quad D^- \\
        Dabar   & D^0\,,\quad \bar{D}^0 \\
	D*0      & D^{*0} \\
        D*+      & D^{*+} \\
        D*-      & D^{*-} \\
        D*0bar   & \bar{D}^{*0} \\
        D*+-     & D^{*+}\,,\quad D^{*-} \\
        D*abar   & D^{*0}\,,\quad \bar{D}^{*0} \\  
	Lc+ &  \Lambda^+_c\\
        Lc*2595+ &  \Lambda^{*+}_c(2595) \\
        Lc*2625+ &  \Lambda^{*+}_c(2625) \\
        Lb0 &  \Lambda^0_b \\
        Lb0bar & \bar{\Lambda}^0_b \\
        Pi0     & \pi^0 \\
        Pi+     & \pi^+ \\
        Nut     & \nu_\tau \\
        Nutbar  & \bar\nu_\tau \\
        Num     & \nu_\mu \\
        Numbar  & \bar\nu_\mu \\
        Nue     & \nu_e \\
        Nuebar  & \bar\nu_e \\
        W+      & W^+ \\
        D**0*0      & D_0^{*0} \\
        D**0*+      & D_0^{*+} \\
        D**0*0bar   & \bar{D}_0^{*0} \\
        D**0*+-     & D_0^{*+}\,,\quad D_0^{*-} \\
        D**0*abar   & D_0^{*0}\,,\quad \bar{D}_0^{*0} \\
        D**1*0      & D_1^{*0} \\
        D**1*+      & D_1^{*+} \\
        D**1*0bar   & \bar{D}_1^{*0} \\
        D**1*+-     & D_1^{*+}\,,\quad D_1^{*-} \\
        D**1*abar   & D_1^{*0}\,,\quad \bar{D}_1^{*0} \\
        D**10      & D_1^{0} \\
        D**1+      & D_1^{+} \\
        D**10bar   & \bar{D}_1^{0} \\
        D**1+-     & D_1^{+}\,,\quad D_1^{-} \\
        D**1abar   & D_1^{0}\,,\quad \bar{D}_1^{0} \\
        D**2*0      & D_2^{*0} \\
        D**2*+      & D_2^{*+} \\
        D**2*0bar   & \bar{D}_2^{*0} \\
        D**2*+-     & D_2^{*+}\,,\quad D_2^{*-} \\
        D**2*abar   & D_2^{*0}\,,\quad \bar{D}_2^{*0}\\
        \hline
        
\end{tabular}
}}
\caption{List of currently available particle specifications and corresponding particles. For each `\textlst{...+}' name, there is a corresponding `\textlst{...-}'.}
\label{tab:particlespecs}
\end{table}

\FloatBarrier

\subsection{Form factor schemes}
\label{sec:FFS}
In general, histogramming of event weights does not commute with contraction of FF parametrization and weight tensors (unless one of the histogram dimensions is explicitly $q^2$).
The \hammer library, therefore, allows the user to specify form factor `schemes' to be used in reweighting. A form factor scheme is a set of FF parameterization choices for
each hadronic transition involving form factors, and is labelled by a `scheme name'. These schemes are set by the method \textlst{Hammer::addFFScheme},
which takes a scheme name plus a map from hadronic string representation to FF parametrization. The hadronic string follows the same syntax and uses the same particle symbols
as for vertex strings in Sec.~\ref{sec:allowforbid}. For example,
\begin{activett}
	ham.addFFScheme("Scheme1", {{"BD", "BLPR"}, {"BD*", "BLPR"}});
	ham.addFFScheme("Scheme2", {{"BD", "BGL"}, {"BD*", "CLN"}});
\end{activett}
declares two different FF schemes, choosing BLPR for both $B \to D$ and $B \to D^*$ form factors in \textlst{"Scheme1"}, and a mixture of schemes for \textlst{"Scheme2"}.
Separate histograms and event weights are generated for each scheme name, which are retrieved with the methods \textlst{Hammer::getHistogram(s)} and \textlst{Hammer::getWeight(s)},
as described below. The list of symbols for available FF parametrizations is provided in Table~\ref{tab:knownampls}. 
The hadronic strings are charge sensitive, so different FFs for charged and neutral processes can be set,
e.g. via an entry \textlst{\{"B+D", "BGL"\}} versus \textlst{\{"B0D", "CLN"\}}, and so on.

Specification of the form factor schemes used to generate the MC sample, i.e., the denominator or input form factors, must be specified in order for \hammer to be able to generate the reweighting tensors. These schemes are specified by the method \textlst{Hammer::setFFInputScheme}, which takes a map from hadronic string representation to FF parametrization scheme. For example
\begin{activett}
	ham.setFFInputScheme({{"BD", "ISGW2"}, {"BD*", "ISGW2"}});
\end{activett}
sets both $B \to D$ and $B \to D^*$ denominator form factors to ISGW2, a common MC parametrization.

As for the include and forbid specifications, the form factor schemes can also be specified in the initialization card in YAML format. The equivalent to the above settings is
\begin{activett}
FormFactors:
	NumeratorSchemes:
		Scheme1: { BD: BLPR, BD*: BLPR }
		Scheme2: { BD: BGL, BD*: CLN }
	Denominator: { BD: ISGW2, BD*: ISGW2 }
\end{activett}

\subsection{Form factor settings}
\label{sec:FFset}
FF parametrization default settings are fixed inside the FF classes themselves. 
Manipulation of the FF default settings may be achieved via \textlst{setOptions}, which takes YAML format arguments. For instance,
\begin{activett}
	ham.setOptions("BtoDBGL: {ChiTmB2: 0.01, ChiL: 0.002}");
\end{activett}
changes the two BGL outer function parameters from their default settings. This can be done before or after invocation of \textlst{initRun}.
Note that the YAML key for the relevant FF class matches the format of the \emph{class prefix}, 
with `\textlst{to}' inserted in the hadronic transition, producing an \textlst{"XtoY"} form. E.g. \textlst{BtoDBGL}, rather than \textlst{BDBGL}.\footnote{
This notation is intended to make it clear we are identifying settings for a particular class -- the $B \to D$ BGL class -- and not a process. 
It further ensures syntactic distinction between a hadronic string representation, which can take a charge assignment like \textlst{"B0D+"}, 
and a class prefix for a $B \to D$ FF class like \textlst{"BtoD"}, which does not. }

See the output of \textlst{Hammer::saveOptionCard} for a full list of the settable form factor parameters and switches, and their default values.

\subsection{Form factor duplication}
\label{sec:FFD}
Duplication of the same FF class is permitted in different FF schemes, and is invoked by adding a token to a FF parametrization name, separated by an underscore. For instance, one may declare
\begin{activett}
	ham.addFFScheme("Scheme1", {{"BD", "BGL_1"}, ... });
	ham.addFFScheme("Scheme2", {{"BD", "BGL_2"}, ... });
	ham.setFFInputScheme({{"BD", "BGL_den"}, ... });
\end{activett}
In this case, three copies of the $B \to D$ BGL class are created, whose settings may be manipulated (via \textlst{setOptions}) separately. E.g.
\begin{activett}
	ham.setOptions("BtoDBGL_1: {ChiT: 0.01, ChiL: 0.002}");
	ham.setOptions("BtoDBGL_2: {ChiT: 0.03, ChiL: 0.007}");
	ham.setOptions("BtoDBGL_den: {ChiT: 0.02, ChiL: 0.005}");
\end{activett}

\subsection{Units}
\label{sec:units}
While the reweights generated by \hammer are dimensionless, various form factor schemes are defined with respect to dimensionful quantities, requiring the library to know the units of the input MC.
This is set by the \textlst{Hammer::setUnits} method, which accepts a string of the name of units convention, from \textlst{eV} to \textlst{TeV}. E.g. \textlst{ham.setUnits("MeV")}, declares the input MC to be in MeV. 
This declaration must be made before \textlst{initRun}. 

The default units inside the library are GeV: The masses and partial widths in the \textlst{Pdg} are specified in GeV. These feed into rate computations, which are therefore also handled internally in GeV. (After events have been processed, \textlst{setUnits} may also be used to specify the units in which partial widths are returned by \textlst{getRate}. See Sec.~\ref{sec:rates} below.) 

\subsection{Processing events}
\label{sec:PE}
An \textlst{Event} object may contain multiple instances of \textlst{Process}, in order to account for the fact that a single event may feature, e.g., two $B$ decay processes.
The \textlst{Event} class is initialized by \textlst{Hammer::initEvent()}, which may take an optional weight double if the event has a non-unit initial weight (this can also be set by \textlst{Hammer::setEventBaseWeight}).
\textlst{Process} instances are added by \textlst{Hammer::addProcess(proc)} which also returns the \textlst{HashId} of the process.
If the process is not allowed according to the \textlst{includeDecay} or \textlst{forbidDecay} specifications, the returned \textlst{HashId} is zero, and the process
is not added to the relevant \textlst{Event} containers.

Once a process is added, it is automatically initialized, which chiefly involves: calculating the signatures of each vertex in the decay cascade; identifying the various subamplitudes making up the cascade,
as well as relevant form factor parametrizations and vertex decay rates, for both the numerator/output and denominator/input; and calculating the total rate for the vertex (this is done only once per run per unique vertex and per FF scheme). The amplitude tensors and form factors are not computed, however,
until the invocation of \textlst{Hammer::processEvent}. Once a process is added, the methods \textlst{Process::getParticlesByVertex} or \textlst{getVertexId} can be used to extract specific
particles in a vertex or other vertex properties, taking as an argument the relevant vertex string. These methods can be used to construct desired observables belonging to the process;
this can also be done by the user externally to \hammer, as desired. E.g.
\begin{activett}
	proc.getParticlesByVertex("D*DPi");
\end{activett}
returns \textlst{pair<Particle, vector<Particle>>} for the parent $D^*$ and vector of daughter particles, $D$ and $\pi$. As an additional convenience, a process can be explicitly
removed from the event by \textlst{Hammer::removeProcess(procId)}, which takes the relevant process \textlst{HashId} as its argument. This functionality is mainly relevant if one
wishes to use \hammer-supplied getter methods for extracting process observables, but one does not actually wish to include the process weight in computations. It can also be used to 
prevent the inclusion of spurious processes in EventIds that would otherwise cause the latter to undesirably proliferate in number.

Once all processes are added (and if histograms have been added, relevant ones have been specified to be filled; see Sec.~\ref{sec:fillhistos}), 
the amplitudes and weight tensors are computed (and weights are added to histogram bins)  
by invocation of \textlst{Hammer::processEvent}. If \textlst{processEvent} is invoked on an event with no included (or all removed) processes, 
\hammer assigns a unit event weight to the event (times any initial weight specified in \textlst{initEvent} or \textlst{setEventBaseWeight}): 
Caution should therefore be employed in invoking \textlst{processEvent} on such events, if this behavior is not desired.

Internally, \textlst{processEvent} proceeds by two separate steps: Calculating the process amplitudes and weight; and then filling histograms (if any).
Either of these steps can be disabled at anytime by the options settings \textlst{ham.setOptions("Hammer: \{CalcProcesses: false\}")} 
and \textlst{ham.setOptions("Hammer: \{CalcHistograms: false\}")}, respectively, and similarly re-enabled at any time. 
Alternatively,  \textlst{processEvent} can accept an optional enum parameter of type \textlst{PAction} as input. 
The default behavior of computing both weights and filling histograms is \textlst{PAction::ALL}, 
but can be modified to weights-only by passing \textlst{PAction::WEIGHTS} or to histograms-only with \textlst{PAction::HISTOGRAMS}. 
In the latter case, the weights must have already been computed in an earlier call.

MC event samples with very large numbers of events but low numerical precision can lead to rare events with `impossible' kinematics. 
For example, in an $X \to YZ$ vertex, if $Y$ and $Z$ have very small angular separation, numerical noise can lead to helicity angle cosines, 
when expressed in terms of kinematic invariants, that fluctuate $> 1$. 
This can lead to a \textlst{NaN} amplitude and event weight. 
The option \textlst{ham.setOptions("ProcessCalc: \{CheckForNaNs: true\}")} allows for explicit checking of \textlst{NaN} at the process amplitude calculational step, 
throwing an error that may be caught upstream as desired (e.g., within a conditional on whether \textlst{processEvent} should fill histograms or take some other action on the problematic event).

\subsection{Setting Wilson Coefficients}
\label{sec:WCFF}
Crucial to the generation of an event weight or histograms 
(via \textlst{Hammer::getHistogram(s)} and \textlst{Hammer::getWeight(s)}, as described below, respectively) 
is setting the relevant `external data', i.e., WCs and FF uncertainties (if any), to be contracted into the relevant tensors.
(Settings for FF parameter central values must be invoked before \textlst{Hammer::processEvent}, as must settings for the denominator/input WCs; see Sec.~\ref{sec:PE}.)  
The WCs are set by the method \textlst{Hammer::setWilsonCoefficients}. The default WC settings are the SM.  A typical example of the usage of this method is
\begin{activett}
	 ham.setWilsonCoefficients("BtoCTauNu", {{"S_qLlL", 1.}, {"T_qLlL",0.25}});
\end{activett}
where the first ``WC space'' argument can be any of \textlst{"BtoCTauNu"}, \textlst{"BtoCTMuNu"}, or \textlst{"BtoCENu"}, 
or  \textlst{"BtoUTauNu"}, \textlst{"BtoUTMuNu"}, or \textlst{"BtoUENu"}. (For $b \to c\ell \nu$ or $b \to u \ell \nu$ processes; 
the valid process arguments are defined by the relevant amplitude classes, and therefore can be different from these.)
If WC specializations are used, then the ``WC space'' argument may take a specialization name suffix, separated by an \textlst{"@"} token; see Sec.~\ref{sec:part_spec_WC}.

The second argument is a \textlst{map<string, complex<double>>} of each WC to its desired value. 
The full list of WCs and their definitions is supplied in Sec.~\ref{sec:NPops}. 
An optional third argument is an enum \textlst{WTerm} value, that declares whether the evaluation should be applied to the numerator and/or denominator (numerator by default). 
The enum \textlst{WTerm} may take values \textlst{ \{COMMON, NUMERATOR, DENOMINATOR\}}
As an alternative, one may instead pass as a second argument
a \textlst{vector<complex<double>>}, corresponding to the ordered basis
\begin{activett}
	{"SM", "S_qLlL", "S_qRlL", "V_qLlL", "V_qRlL", "T_qLlL",
				   "S_qLlR", "S_qRlR", "V_qLlR", "V_qRlR", "T_qRlR"}.
\end{activett}

It is important to note that the \textlst{setWilsonCoefficients} method, when taking a \textlst{map}, produces \emph{incremental} settings changes. 
I.e. the sequential invocations
\begin{activett}
	ham.setWilsonCoefficients("BtoCTauNu",
	 				{{"S_qLlL", 1.}, {"T_qLlL",0.25}});
	ham.setWilsonCoefficients("BtoCTauNu", {{"S_qLlL", 0.5}});
\end{activett}
will result in \textlst{S\_qLlL} $= 0.5$ and \textlst{T\_qLlL} $= 0.25$, since the latter was not affected by the second call.
The method \textlst{resetWilsonCoefficients} takes the WC type -- e.g. \textlst{"BtoCTauNu"}-- and resets the corresponding WCs to the default SM. 
One may check what the current state of the WC settings are with \textlst{retrieveWilsonCoefficients} 
which takes a ``WC space'' argument and optional \textlst{WTerm} enum value \textlst{NUMERATOR} or \textlst{DENOMINATOR},
and returns a \textlst{map<string, complex<double>>}.

\subsection{Setting FF eigenvectors}
\label{sec:set_FF_eigs}
As mentioned in Sec~\ref{sec:HG}, certain form factor classes (typically, those with names ending in ``\textlst{Var}'') incorporate linearized variation of the FF parametrization, 
with additional variational indices in the form factor tensors, in the sense defined by eq.~\eqref{eqn:FFerr}. 
This generalizes the tensor weights into the form factor error eigenspace (or whatever space is defined in the relevant parametrization's class), 
which may then be contracted with the desired error (eigen)vector, permitting reweighting of the events to any point in this space. 

This error (eigen)vector is set via the method \textlst{Hammer::setFFEigenvectors}. 
The usage is similar \textlst{setWilsonCoefficients}, except that \textlst{setFFEigenvectors} takes the name of the hadronic process in \textlst{"XtoY"} form (see Sec.~\ref{sec:FFS}), 
the name of the FF parametrization, and then either a \textlst{map<string, double>} of the error coordinates to be changed, or a vector of coordinates \textlst{vector<double>}, 
with respect to the basis defined by the parametrization's class. 
A typical example of the usage of this method is
\begin{activett}
	 ham.setFFEigenvectors("BtoD*", "BGLVar", {{"delta_a1", 0.1}, {"delta_b1",-0.05}});
\end{activett}
See Sec.~\ref{sec:FFUC} for examples of definitions and conventions of currently implemented FF variational classes. 
Parametrizations with FF uncertainty indices are intended to be used only in the numerator FF schemes; 
specific settings for the denominator classes can be implemented by a duplicated FF scheme and \textlst{setOptions} (see Sec.~\ref{sec:FFS}).

The FF classes with linearized variations permit the matrix of eigenvectors of the fit covariance -- the eigenspace matrix, which defines the basis of variations -- 
to be set as an option through \textlst{setOptions}, e.g. if the eigenspace matrix setting is called \textlst{"<eigmatrix>"} in the relevant class, then
\begin{activett}
	ham.setOptions("BtoDBGLVar: {<eigmatrix>: [[v11,v12,...],[v21,v22,...],...]}");
\end{activett}
(The special choice that this eigenspace matrix is the identity typically corresponds to the choice 
that each eigendirection is actually just motion in the underlying linearized space of FF parameters.)
These classes similarly permit the naming scheme for the basis of variations to be adjusted by changing the vector of names using the method \textlst{Hammer::renameFFEigenvectors},
in order to accommodate different conventions for this matrix. 
For example, if the eigenspace matrix is the identity, it is clearer to label the basis of variations with respect to the parameter names, 
e.g. \textlst{"delta_a1"} for variation in the $a_1$ $B \to D^*$ BGL parameter and so forth (see Sec.~\ref{sec:FFUC}).
However, if the eigenspace matrix is instead the actual eigenvectors, one might prefer \textlst{"delta_e1"} and so on.

For example, if the $X \to Y$ FF class \textlst{DemoVar} has default basis \textlst{\{delta_a, delta_b, delta_c, delta_d\}}, this may be changed via
\begin{activett}
	 ham.renameFFEigenvectors("XtoY", "DemoVar", {"delta_e1", "delta_e2", "delta_e3", "delta_e4"});
\end{activett}
This can be done before or after invocation of \textlst{initRun}. 
A warning will be thrown if the renaming list is longer than the basis defined in the class. 
Passing a list of $k$ names that is shorter than defined in the class will rename only the first $k$; an empty string at the i-th position leaves the name at the i-th position unchanged.
For example, the list of names \textlst{\{"", "delta_e2"\}} in the above example would leave \textlst{"delta_a"}, change \textlst{"delta_b"}, then leave all remaining names unchanged.

As for WCs, the \textlst{setFFEigenvectors} method, when taking a \textlst{map}, produces \emph{incremental} settings changes.
The method \textlst{resetFFEigenvectors} takes the name of the hadronic process in \textlst{"XtoY"} form and the name of the FF parametrization, 
and resets the corresponding FF eigenvectors to zero.
One may check what the current state of the FF eigenvector settings are with \textlst{retrieveFFEigenvectors} 
which takes the hadronic process in \textlst{"XtoY"} form and the name of the FF parametrization
and returns a \textlst{map<string, double>}.

\subsection{(Partial) specialization of Wilson Coefficients}
\label{sec:part_spec_WC}
The default $11$-dimensional WC space for a specific process (see Sec.~\ref{sec:WCFF}) can be projected into arbitrary subspaces via the use of a \emph{partial specialization}. 
This is of particular utility when, e.g., fitting various simplified models, in which only a small set of linear combinations of WCs are needed, 
leading to significant speed enhancements and memory requirement reductions. 
One may also specify a full specialization, equivalent to fixing the WCs of a process set to a zero-dimensional subspace, i.e. a fixed set of values.

A specialization is defined on a particular WC space (e.g. \textlst{"BtoCTauNu"}), and consists of: 
a name, a list of the new coordinate names for the WC subspace, an origin, and a definition of the basis vectors corresponding to the new coordinates. 
Because specializations may be applied to event weights or histograms (see Sec.~\ref{sec:histo_part_spec} below) or both, 
the creation, definition, and application of a specialization is handled through separate methods.

Creation of a specialization is achieved through \textlst{Hammer::createWCSpecialization}, which takes a specialization name, a WC space, and a list of coordinate names.
For example
\begin{activett}
	ham.createWCSpecialization("JM", "BtoCTauNu", {"c1", "c2", "c3"});
\end{activett}
creates a partial specialization called \textlst{"JM"} in a three-dimensional subspace with coordinates \textlst{"c1"}, \textlst{"c2"}, and  \textlst{"c3"}. 
This method must be applied before \textlst{initRun}.
A full specialization corresponds to a zero-dimensional subspace, and thus the list of coordinates for such a specialization can be passed as empty, or omitted entirely.

An arbitrary number of specializations can be created for any given WC space
A specialization by the same name can be created for other WC spaces, with entirely different properties (coordinates, origin, basis). For example, one could add
\begin{activett}
	ham.createWCSpecialization("JM", "BtoCMuNu", {"x1", "x2"});
\end{activett}
that would act on $b \to c \mu \nu$ processes. 
In this respect, a specialization should be thought of as the WC equivalent of an FF scheme, 
defining a set of weights (or histograms) computed and retrievable under it for each type of process in the event sample being reweighted.

By default, \textlst{createWCSpecialization} also defines a subspace origin corresponding to the current settings set by \textlst{setWilsonCoefficients} 
(if uninvoked, the WC settings default is the SM). 
This origin may be explicitly set via \textlst{Hammer::setWCSpecializationOrigin}, which takes a specialization name, a WC process set, 
and coordinates specified either as a map or vector of values (in the original WC basis). 
For example,
\begin{activett}
	ham.setWCSpecializationOrigin("JM", "BtoCTauNu", {{"SM", 1}, {"V_qRlL", 0.5}});
\end{activett}
fixes the origin of the subspace at the SM, plus a contribution from a right-handed vector current. This method must be applied before \textlst{initRun}.

For partial specializations, one must also define the basis of the subspace, which may be done either coordinate by coordinate 
via \textlst{Hammer::setWCSpecializationCoord} or instead \textlst{Hammer::setWCSpecializationBasis}. 
The former takes a specialization name, a WC space, a coordinate name, 
plus a map or vector of values that defines the new coordinate basis vector in terms of a linear combination of the original WC basis vectors.
The latter takes instead a specialization name, a WC space, and a vector of maps, 
each of which defines a new coordinate basis vector in terms of a linear combination of the original WC basis vectors, 
following the order of coordinates defined in \textlst{createWCSpecialization}.
For instance,
\begin{activett}
	ham.setWCSpecializationCoord("JM", "BtoCTauNu", "c1", 
					 {{"S_qRlL", 1}, {"T_qLlL", 0.25}});
	ham.setWCSpecializationCoord("JM", "BtoCTauNu", "c2",  
					{{"V_qRlL", 1}, {"V_qLlL", 1}});
	ham.setWCSpecializationCoord("JM", "BtoCTauNu", "c3",  
					{0,0,2,0,0,0,0,0,0,0,0});
\end{activett}
is the same as
\begin{activett}
	ham.setWCSpecializationBasis("JM", "BtoCTauNu", { 
		{{"S_qRlL", 1}, {"T_qLlL", 0.25}},
		{{"V_qRlL", 1}, {"V_qLlL", 1}},
		{{"S_qRlL", 2}}});
\end{activett}
These are equivalent to defining three new coordinates $c_{1,2,3}$ to have basis vectors $e_1 = S_{qRlL} + 0.25 T_{qLlL}$, $e_2 = V_{qRlL} + V_{qLlL}$, and $e_3 = 2S_{qRlL}$.
These methods must be applied before \textlst{initRun}.

Once fully specified, the application of the specialization---whether to weights or histograms or both---must 
be specified respectively by \textlst{Hammer::applyWCSpecializationInWeights}, \textlst{Hammer::applyWCSpecializationInHistogram} 
or \textlst{Hammer::applyWCSpecializationInAllHistograms}.
The latter two are discussed in Sec.~\ref{sec:histo_part_spec} below.
The former takes just the name of the specialization. 
Once invoked, \textlst{processEvent} will only compute weights with this (and any other) specializations applied,
and computation of the `general weights', i.e., the tensor weights in the original WC space with no specialization, is disabled,
and they will therefore not be available to be saved to buffers (see Sec.~\ref{sec:buf}).
One may require that general weights are to be computed and save-able by \textlst{saveEventWeights} via
\begin{activett}
	ham.saveEventGeneralWeights(true);
\end{activett}	
One may also reenable only the computation of the general weights via the option setting
\begin{activett}
	ham.setOptions("Hammer: {CalcGeneralWeights: true}");
\end{activett}
When histograms are intended to be generated with different (or no) specializations, 
the \textlst{applyWCSpecializationInWeights} method should be applied before \textlst{initRun}, 
or otherwise \textlst{reconcileSpecializations()} may need to be subsequently run 
to ensure the weight tensors required for histograms are generated.

For each specialization applied to weights, the corresponding specialized rate is also computed, along with the `general rate', 
i.e., the tensor rate in the original WC space with no specialization.
See Sec.~\ref{sec:rates}.
The general rate computation (and subsequent saving, see Sec.~\ref{sec:saveheadevrate}) can be disabled by 
\begin{activett}
	ham.setOptions("Hammer: {GeneralRates: false}");
\end{activett}

A specialization may be removed via \textlst{Hammer::removeWCSpecialization}, which takes the specialization name 
and an optional bool to remove computed event weight or histogram data (false by default).
This method removes the specialization on all WC spaces with the given name. 
The method \textlst{Hammer::removeAllWCSpecializations} removes all extant specializations, taking the same optional bool.
The application of a specialization to just the weight calculation can be disabled by \textlst{Hammer::removeWCSpecializationInWeights}, 
which takes the specialization name.  
If histograms are being generated, \textlst{reconcileSpecializations()} may need to be run after these removal methods, see Sec.~\ref{sec:histo_part_spec}.

The names of all currently-created specializations can be obtained from the method \textlst{availableWCSpecializationIds()},
and the list of specializations that are currently applied to the event weight calculation can be obtained from \textlst{appliedWCSpecializationsInWeights()}.

The coordinates of the WC specialization subspace may be set using \textlst{setWilsonCoefficients}, 
by modifying the name of the original WC space to contain a specialization name suffix, separated by an \textlst{"@"} token. For example,
\begin{activett}
	ham.setWilsonCoefficients("BtoCTauNu@JM", {{"c1", 1}});
	ham.setWilsonCoefficients("BtoCTauNu@JM", {1,0,0});
\end{activett}
are equivalent means to set the first coordinate of the \textlst{"JM"} specialization of the \textlst{"BtoCTauNu"} space.

\subsection{Retrieving event weights}
Once an event has been processed (or loaded from a file), the weight for the currently loaded event can be retrieved via \textlst{Hammer::getWeight("FFScheme")}.
The method  \textlst{Hammer::getWeights("FFScheme")}, 
instead returns a map of each process Id and corresponding \textlst{double} process weight for the specified FF scheme. 
These weights can then be combined as appropriate. Alternatively,  
if one wants weights for a restricted set of processes, identified by \textlst{HashId}'s, 
one may use \textlst{Hammer::getWeight("FFScheme", procIds)}, where the second argument is a \textlst{vector<HashId>},
that returns the corresponding weights already combined into a \textlst{double}.

For the case that a WC specialization has been applied to the weight calculation (see Sec.~\ref{sec:part_spec_WC}), 
all these methods take an optional last argument specifying the specialization name 
(the default is \textlst{""}, corresponding to returning the general weight, if it has been computed; see Sec.~\ref{sec:part_spec_WC} and~\ref{sec:histo_part_spec}).

\subsection{Histograms}
\label{sec:histos}
Histograms of arbitrary dimensionality may be created by the \hammer library. In general, histogram bins contain event weight tensors, which are \emph{direct products}
of the process weight tensors for all processes in the event that are included by an \textlst{includeDecay} specification (and not specifically removed by a later \textlst{removeProcess} invocation).  It is up to the user to determine programmatically which processes in an event are (or are not) included. For example, under the include specification shown in Sec.~\ref{sec:allowforbid}, an event featuring $\bar{B}^0 \to (D^{*+} \to (D^+ \to K^+ \pi^+ \pi^-)\gamma)(\tau^- \to \ell^-\nu\nu)$ and $B^0 \to D^{-}\mu^+\nu$ would have an event weight composed from the product of both process weights, while an event featuring $\bar{B}^0 \to (D^{+}(\tau^- \to \ell^-\nu\nu)$ and $B^0 \to D^{-}\mu^+\nu$ would just have an event weight equal to the process weight for the $B^0 \to D^{-}\mu^+\nu$ decay.

The event weight tensor may be contracted with arbitrary WCs to generate \emph{a posteriori} the corresponding histogram bin weight. Thus, once a histogram is computed,
it is computed for all NP. More specifically, a contracted histogram contains elements that are \textlst{BinContents} structs, 
with members \textlst{sumWi}, \textlst{sumWi2}, and \textlst{n} for sum of weights, sum of squared weights and number of events in the bin, respectively.

\subsubsection{Adding}
\label{sec:addhistos}
A histogram is declared by \textlst{Hammer::addHistogram}, which takes as arguments a name string and either: a vector of dimensions, 
a bool for under/overflow and a vector of ranges; or a vector of bin edges and a bool for under/overflow.
The method \textlst{addHistogram} does not create a single histogram, but rather a \emph{histogram set}: 
A separate histogram is created for each unique \emph{event ID} and, in turn, for each FF scheme name specified by \textlst{addFFScheme}. 
Here, an event ID is a \textlst{set} of process IDs for all processes included in the event.\footnote{
Because of histogram compression functionality discussed in Sec.~\ref{sec:compresshistos} below, histograms are in practice indexed by a \emph{event ID group}, 
which is a set of event IDs: Without compression, each event ID group is just a trivial single element set containing the event ID of the histogram.}
For instance
\begin{activett}
	ham.addHistogram("q2VsEmu", {20, 15}, false, {{3.,12.},{0,2.5}});
\end{activett}
creates a \emph{histogram set} each with $20\times 15$ bins, no under/overflow, binned uniformly over the respective ranges $3$--$12$ and $0$--$2.5$ (in appropriate units). 
With reference to the above \textlst{addFFScheme} example in Sec.~\ref{sec:FFS}, 
this histogram set contains one histogram for each combination of either \textlst{"Scheme1"} or \textlst{"Scheme2"} with each unique $B \to D$ decay cascade. 
 Alternatively, for non-uniform bins
\begin{activett}
	ham.addHistogram("q2VsEmu", {{3.,5.,9.,12.},{0,1,2.5}}, true);
\end{activett}
which creates a $3\times2$ histogram, with additional under/overflow bins.
For an MC sample with $n$ unique event IDs and $m$ declared FF schemes, the former  \textlst{addHistogram} invocation would create $m\times n$ unique $20 \times 15$ histograms, all with the name \textlst{"q2VsEmu"}.  

\subsubsection{Filling}
\label{sec:fillhistos}
Filling of histograms for a specific event is performed by \textlst{Hammer::fillEventHistogram}, which takes the histogram name and the values of the observables corresponding to each histogram dimension. (A deprecated method 
\textlst{Hammer::setEventHistogramBin} takes the indices of the bin to be filled.) For example,
\begin{activett}
	ham.fillEventHistogram("q2VsEmu", {4., 0.5});
\end{activett}
fills the appropriate bin element for the \textlst{"q2VsEmu"} histograms belonging to the event being processed, and fills the relevant histograms for each FF scheme name.  Invocations of \textlst{fillEventHistogram} must occur before \textlst{Hammer::processEvent}. Otherwise, the relevant histogram will not be filled with the weight for event being processed: If \textlst{fillEventHistogram} is not invoked for a particular histogram for a particular event, the event weight is not added to the histogram. When the under/overflow bool is set to \textlst{false}, events outside the bin ranges are ignored by \textlst{fillEventHistogram}.

A single bin histogram set \textlst{"Total Sum of Weights"} may be created via the method \textlst{Hammer::addTotalSumOfWeights}, which takes additional bools for collapsing processes and uncertainties (see Sec.~\ref{sec:histo_part_spec}). This method should invoked before \textlst{initRun}. The \textlst{"Total Sum of Weights"} histogram, if it has been created, is automatically filled by \textlst{processEvent}. 

\subsubsection{Compression}
\label{sec:compresshistos}
In many use cases, the entire histogram set is not required, but rather its direct sum. Computing and storing only the latter compressed form permits both speed gains and space savings. The method \textlst{Hammer::collapseProcessesInHistogram} takes a name of a histogram, and causes all members of the histogram set containing the same tensor structures to be summed and collapsed into a single compressed histogram. For instance,
\begin{activett}
	ham.collapseProcessesInHistogram("q2VsEmu");
\end{activett}	
This method should invoked before \textlst{initRun}. 
When invoked, each compressed histogram in the histogram set is then indexed by non-trivial event ID groups, containing the event IDs of all the histograms that were collapsed into it.

\subsubsection{(Partial) specialization}
\label{sec:histo_part_spec}
A declared WC specialization (see Sec.~\ref{sec:part_spec_WC}) may be applied to a particular histogram via \textlst{Hammer::applyWCSpecializationInHistogram}, 
which takes the name of the histogram and the name of the specialization as arguments. 
For example,
\begin{activett}
	ham.applyWCSpecializationInHistogram("q2VsEmu", "JM");
\end{activett}
This method should be applied before \textlst{initRun}, 
as the \textlst{initRun} method reconciles the various applications of specializations to the weight calculation and histogram generation, 
in order to determine whether general weights need to be computed.
If one is applying or removing specializations after the declaration of \textlst{initRun}, 
one may explicitly reinvoke this reconciliation procedure via \textlst{Hammer::reconcileSpecializations()}.
One may also apply a specialization to all known histograms at once via
\begin{activett}
	ham.applyWCSpecializationInAllHistograms("JM");
\end{activett}

When \textlst{applyWCSpecializationInHistogram} is invoked, the generation of the histogram containing the general weight tensors, 
i.e. tensor weights in the original WC space with no specialization, is disabled.
This may be explicitly re-enabled for histogram \textlst{"<name>"} via
\begin{activett}
	ham.setOptions("Histos: {CalcGeneralHistogram_<name>: true}");
\end{activett}

In addition to WC specialization, for a specific histogram one may wish to fix \emph{a priori} all FF variational indices for a particular scheme, 
in order to reduce space or reweighting times.
This is achieved with the method \textlst{specializeFFInHistogram} that takes the histogram name plus the arguments required by \textlst{setFFEigenvectors}.
As an example,
\begin{activett}
	ham.specializeFFInHistogram("q2VsEmu", "BtoD*", "BGLVar", {{"delta_a1", 0.1}, {"delta_b1",-0.05}});
\end{activett}
or alternatively the last argument may be a vector of values with respect to the basis defined by the FF parametrization’s class.
This method should be invoked \emph{after} \textlst{initRun}.
It is the FF equivalent to a full specialization of a WC space. 
We do not provide partial specialization syntax for FF variational classes in the same way as for WCs, 
as this instead can be simply achieved through redefinition of the rank of the eigenspace matrix using \textlst{setOptions} (see Sec.~\ref{sec:set_FF_eigs}),
combined with the already-existing FF duplication functionalities in the FF scheme declarations.
FF specialization is not reversible once a histogram is filled in an initialization run. 

The reset method \textlst{Hammer::removeWCSpecializationInHistogram}, taking the histogram name and specialization, is also provided,
which may e.g. be invoked before a subsequent initialization run to reset the histogram definition to the default and turn off its WC specializations.
Similarly, \textlst{Hammer::removeFFSpecializationInHistogram}, taking just the histogram name, can deactivate an FF specialization for a future initialization run.
When these methods are applied, it may be necessary to run the specialization reconciliation procedure via \textlst{Hammer::reconcileSpecializations()}.

When merging different samples of previously computed histograms (see Sec.~\ref{sec:parallel_merge}),
the library will merge histograms (per unique Event ID and FF scheme) that have been specialized with the \emph{same WC specialization name}. 
Care should be taken to ensure programmatic consistency.

\subsubsection{Retrieval}
\label{sec:retrievehistos}
Once all events or histograms have been processed (or reloaded from a file, see Sec.~\ref{sec:buf}) the user may retrieve a specific histogram via the method \textlst{Hammer::getHistogram}, which takes a histogram name and a FF scheme name.  NP choices must be specified first via \textlst{setWilsonCoefficients}, as must FF uncertainties via \textlst{setFFEigenvectors} if a parametrization in the desired FF scheme has them. For example,
\begin{activett}
	ham.setWilsonCoefficients("BtoCTauNu", {{"S_qRlL", 1.},{"S_qLlL", 0.5}});
	auto histo = ham.getHistogram("q2VsEmu", "Scheme2");
\end{activett}
would contract the bin weights with the specified NP Wilson coefficients (and FF eigenvectors, if any) for each histogram in the \textlst{"q2VsEmu"} histogram set populated for \textlst{"Scheme2"}, and then combine them together into a single final histogram.
This contracted histogram output \textlst{histo} is a (row-major) flattened vector of \textlst{BinContents} structs.
By contrast, the method \textlst{getHistograms} (note the plural) extracts all histograms of a specific name and scheme. For example,
\begin{activett}
	auto histos = ham.getHistograms("q2VsEmu", "Scheme2");
\end{activett}
produces a map of eventIDs to histogram for all available \textlst{"q2VsEmu"} histograms with FF scheme \textlst{"Scheme2"}.
If \hammer has been compiled with \texttt{ROOT} support enabled, then the \texttt{ROOT}-specific methods \textlst{Hammer::getHistogram1D|!{\CPPidentifierstyle{2D}}\lstinline! |!{\CPPidentifierstyle{3D}}\lstinline!} and \textlst{Hammer::getHistograms1D|!{\CPPidentifierstyle{2D}}\lstinline! |!{\CPPidentifierstyle{3D}}\lstinline!} will be available. They allow, for the case of 1D, 2D, and 3D histograms, to obtain the contracted histogram(s) as \texttt{ROOT}'s \textlst{TH1D|!{\CPPidentifierstyle{2D}}\lstinline! |!{\CPPidentifierstyle{3D}}\lstinline!} objects respectively. These methods have the same syntax as the corresponding non-\texttt{ROOT} ones.

\subsubsection{Uncertainties}

Computation of the weight-squared uncertainties (accessed from the \textlst{BinContents} struct via \textlst{sumWi2}) is off by default. This may be enabled globally via the options setting \textlst{ham.setOptions("Histos: \{KeepErrors: true\}")}. However, for computational speed and/or memory efficiency, it may be instead enabled or disabled for individual histograms via \textlst{Hammer::keepErrorsInHistogram}, which takes the name of the histogram as an argument, and a bool. For instance
\begin{activett}
	ham.keepErrorsInHistogram("q2VsEmu", true);
\end{activett}
enables weight-squared computation for this particular histogram. This method should be invoked before \textlst{initRun}.

\subsubsection{Projection}
On occasion, it may be useful to project a pre-computed $n$-dimensional histogram onto a lower-dimensional one. This can be achieved via the method \textlst{createProjectedHistogram}, which takes the name of the original $n$-dimensional histogram, the name of the new histogram to be created, and a set of the index positions to be summed over or collapsed. For instance, for a $3$-dimensional histogram \textlst{"q2VsEmuVsM2miss"} with dimensions $q^2$, $E_{\mu}$ and $m^2_{\text{miss}}$, one may integrate over the $E_\mu$ and $m^2_{\text{miss}}$ dimensions to create a $1$-dimension $q^2$ histogram via
\begin{activett}
	ham.createProjectedHistogram("q2VsEmuVsM2miss", "justq2", {1,2});
\end{activett}
in which the new histogram, named \textlst{"justq2"}, inherits the underlying structure -- the histogram set -- of the original histogram.

\subsection{Pure phase space vertices}
\label{sec:purePS}
The \hammer library permits the user to declare particular vertices, in either the denominator or numerator amplitude, to be evaluated as pure phase space. This is achieved by the method \textlst{Hammer::addPurePSVertices}, which takes a set of string vertices as an argument, and an optional enum \textlst{WTerm} value to declare whether the evaluation should be applied to the numerator and/or denominator (numerator by default). The enum \textlst{WTerm} has values \textlst{ \{COMMON, NUMERATOR, DENOMINATOR\}}

As an example
\begin{activett}
	ham.addPurePSVertices({"TauMuNuNu","D*+DPi"});
	ham.addPurePSVertices({"D*DGamma"}, WTerm::DENOMINATOR);
\end{activett}
declares all $\tau \to \mu \nu \nu$ and $D^{*+}\to D\pi$ vertices in the numerator and all $D^* \to D\gamma$ vertices in the denominator, to be evaluated as phase space 
(subject to the rules below). The equivalent initialization card definition is
\begin{activett}
PurePSVertices:
	Numerator: [ TauMuNuNu, D*+DPi ]
	Denominator: [ D*DGamma ]
\end{activett}

The library employs the pure phase space definition
\begin{equation}
	\label{eqn:PSdef}
	 \frac{1}{\prod_k |\{s_k\}|} \sum_{s_i, r_j} \big|\mathcal{M}_{s_1,\ldots, s_n; r_1,\ldots,r_m}\big|^2 = 1 \times (m^{6-2n})\,,
\end{equation}
where $s_i$ ($r_i$) are incoming (outgoing) quantum numbers, $|\{s_k\}|$ is the number of states of $s_k$, $m$ is the mass of the parent particle in the vertex, and $n$ the number of daughters. I.e., the squared matrix element averaged over initial states and summed over final states is set to unity times a factor that preserves the dimensionality of the overall amplitude. Upon the declaration of a vertex as PS, averaging over the initial states of all immediate (non-PS) daughter vertices is automatically performed.

The declaration of a vertex as phase space within an edge may be ambiguous, if the other vertex is not declared as PS too. This ambiguity is resolved by the library by an \emph{exclusive} implementation of the \textlst{addPurePSVertices} method, according to the following rules:
\begin{enumerate}
	\item If both vertices in an edge are declared as PS, the edge is set to PS.
	\item The declaration of a single vertex in an edge as PS is obeyed only if the remaining vertex has a known vertex amplitude.
\end{enumerate}
Labelling a PS declaration by an underlaid cross, i.e. \cvertex[ps] or \overtex[ps], these rules are represented as follows:
\begin{center}
\newcolumntype{L}{ >{\raggedright\arraybackslash} m{4cm} <{}}
\begin{tabular}{Ll}
	\puredge[ps][ps] 	& Edge is set to PS  \\
	\puredge[ps] 		& Declaration refused; a warning is thrown \\
	\paredge[ps][ps]	& Edge is set to PS \\
	\paredge[ps][]		& Declaration refused; a warning is thrown \\
	\paredge[][ps]		& Edge is replaced by remaining \cvertex \\
	\fuledge[ps][ps]		& Edge is set to PS \\
	\fuledge[ps]		& Edge is replaced by remaining \cvertex \\
\end{tabular}
\end{center}
An example of these rules is shown in Table~\ref{tab:exampleps} for the examples of Table~\ref{tab:examplearith}, based on the process tree in Fig.~\ref{fig:processtree}. In the first example, the declaration of vertex 1 as pure phase space is accepted, as the $0$--$1$ edge is replaced by the known vertex amplitude at vertex $0$. In the second example, the declaration is refused, since vertex $5$ cannot be evaluated independently.

\begin{table}[h]
\newcolumntype{C}{ >{\centering\arraybackslash} m{5cm} <{}}
\begin{tabular}{C|C}
\hline
Known Amplitudes & Evaluated Amplitudes \\
\hline\hline
\vspace{5pt}
\begin{tikzpicture}
		\tikzset{every node/.style={scale = 0.7, transform shape}}
		\tikzstyle{level 1}=[sibling angle=50, level distance = 0.7*2cm];
		\tikzstyle{level 2}=[sibling angle=50, level distance = 0.7*2cm];
		\tikzstyle{level 3}=[sibling angle=50, level distance = 0.7*2cm];
		\tikzstyle{o}=[draw=black!80,line width=1.5pt,fill=white,circle,inner sep=2pt,minimum size = 12pt,align=center, text=black]
		\tikzstyle{ops}=[draw=black!80,line width=1.5pt,fill=white,circleps,inner sep=2pt,minimum size = 12pt,align=center, text=black]
		\tikzstyle{c}=[draw=black!80,line width=1.5pt,fill=black!80,circle,inner sep=2pt,minimum size = 12pt,align=center, text=white]
		\tikzstyle{edge from parent}=[draw=black!40, line width = 2pt]
		\node (P) [c] {0} [grow cyclic,shape=circle, clockwise from=-65]
			child  { [grow cyclic,shape=circle, clockwise from=-65] node [ops] {1} edge from parent [draw = black!80, line width = 1.5pt, double]
				child  {node [c] {5} edge from parent [draw = black!80, line width = 2pt]
				}
			}
			child  {node [c] {2} edge from parent [draw = black!80, line width = 2pt]
			};
\end{tikzpicture}
& \vvertex{0}, \vvertex{2}, \vvertex{5} \\
\hline
\vspace{5pt}
\begin{tikzpicture}
		\tikzset{every node/.style={scale = 0.7, transform shape}}
		\tikzstyle{level 1}=[sibling angle=50, level distance = 0.7*2cm];
		\tikzstyle{level 2}=[sibling angle=50, level distance = 0.7*2cm];
		\tikzstyle{level 3}=[sibling angle=50, level distance = 0.7*2cm];
		\tikzstyle{o}=[draw=black!80,line width=1.5pt,fill=white,circle,inner sep=2pt,minimum size = 12pt,align=center, text=black]
		\tikzstyle{ops}=[draw=black!80,line width=1.5pt,fill=white,circleps,inner sep=2pt,minimum size = 12pt,align=center, text=black]
		\tikzstyle{c}=[draw=black!80,line width=1.5pt,fill=black!80,circle,inner sep=2pt,minimum size = 12pt,align=center, text=white]
		\tikzstyle{edge from parent}=[draw=black!40, line width = 2pt]
		\node (P) [c] {0} [grow cyclic,shape=circle,clockwise from=-65]
			child  { [grow cyclic,shape=circle,clockwise from=-65] node [ops] {1} edge from parent [draw = black!80, line width = 1.5pt, double]
				child  {node [o] {5} edge from parent [draw = black!80, line width = 1.5pt, double]
				}
			}
			child  {node [c] {2} edge from parent [draw = black!80, line width = 2pt]
			};
\end{tikzpicture}
&  \vvertex{0}, \vedge{1}{5}, \vvertex{2} \\
\hline
\end{tabular}
\caption{Example arithmetic for filling amplitudes for the examples of Tab.~\ref{tab:examplearith}, with an additional phase space declaration on vertex 1.}
\label{tab:exampleps}
\end{table}

\subsection{PHOTOS}
\label{sec:photos}
Typical MC samples include soft and collinear radiative corrections, incoherently appended to the relevant vertices by the \texttt{PHOTOS} algorithm~\cite{Photos:1991ph}, ignoring typically negligible interference effects. Inclusion of such (typically very soft) radiative photons requires the vertex (and all daughter vertex) momenta to be rebalanced, such that overall momentum remains conserved. For the purpose of reweighting the truth level process, these photons must be pruned from the process tree, which in turn requires a reversion of the kinematic rebalancing. (As such, because they are automatically pruned, radiative photons need not be specified in \textlst{includeDecay} or \textlst{forbidDecay} specifications.)

The effect of the kinematic rebalancing on the actual event weight is generally negligible: The main concern is to ensure momentum conservation in the process tree once the photon is removed. With this in mind, and following the \texttt{PHOTOS} prescription for kinematic rebalancing~\cite{Photos:1991ph}, the \hammer library therefore identifies radiative photons, and reverts the kinematics to pre-radiative corrected form, by the following procedure:
\begin{enumerate}[noitemsep, topsep =0pt] 
	\item If a vertex contains 3 or more particles, with at least one photon, the softest photon is identified as radiative.
	\item  The radiative photon, $\gamma_{\text{rad}}$, is assumed to be associated with the nearest charged particle, labelled `ch', in the polar angle distance, $\delta \theta$.
	\item The radiative vertex, and all daughter particles, are then partitioned into: The parent particle, `P'; The charged particle, `ch', and all its descendants, the `ch subtree'; All other particles in the radiative vertex except $\gamma_{\text{rad}}$, collectively called `$Y$', and all their descendants, the `$Y$ subtree'. The radiative vertex is thus written $P \to \text{ch} + Y + \gamma_\text{rad}$.
	\item The ch and $Y$ subtrees are boosted to the $p_{\text{ch}}+p_Y$ rest frame, $R_{\text{ch} + Y}$, so that necessarily $\bm{p}_{\text{ch}}$ and $\bm{p}_Y$ are back-to-back.
	\item In $R_{\text{ch} + Y}$ frame, writing $p_Y = (E_Y, \bm{p}_Y)$ and $p_\text{ch} = (E_\text{ch}, \bm{p}_\text{ch})$, the ch subtree and Y subtree are then \emph{independently} longitudinally boosted by
	\begin{equation}
		\beta\gamma_\text{ch} = \frac{E_{\text{ch}} |\bm{p}^*| - E_{\text{ch}}^* |\bm{p}_{\text{ch}}|}{m_{\text{ch}}^2}\,,\qquad \beta\gamma_Y = \frac{E_Y |\bm{p}^*| - E_Y^* |\bm{p}_Y|}{m_Y^2}\,,
	\end{equation}
	in which the starred quantities are the usual $P$ rest frame kinematic objects for the two body decay $P \to \text{ch} + Y$, i.e.
	\begin{equation}
		E_{\text{ch}}^* = \frac{m_P^2 - m_Y^2 + m_\text{ch}^2}{2m_P}\,,\qquad E_Y^* = \frac{m_P^2  -  m_\text{ch}^2 + m_Y^2}{2m_P}\,,\qquad |\bm{p}^*| = \frac{m_P}{2}\lambda^{1/2}\bigg[\frac{m_Y}{m_P}, \frac{m_{\text{ch}}}{m_P}\bigg]\,,
	\end{equation}
	with $\lambda(x,y) = (1 - (x+y)^2)(1-(x-y)^2)$. Under these independent boosts, momentum conservation is restored to the $P \to \text{ch} + Y$ vertex with $\gamma_\text{rad}$ removed.
	\item The ch and $Y$ subtrees are then boosted to the frame such that $p_{\text{ch}} + p_{Y} = p_P$, the latter meaning the actual momentum of particle $P$ in the process tree.
	\item This process is repeated until (i) is no longer true.
\end{enumerate}

\subsection{Rates}
\label{sec:rates}
The library provides the means to compute the partial width for a particular vertex via \textlst{Hammer::getRate}, 
which takes as argument either a vertex string or the parent and daughter PDG codes, 
plus a scheme and an optional WC specialization (see Sec.~\ref{sec:part_spec_WC}).
It may also take a vertex hash ID, obtainable from a specific process via \textlst{Process::getVertexId}. 
Partial widths are returned in the units specified by \textlst{Hammer::setUnits}; the default is GeV (see Sec.~\ref{sec:units}). For example 
\begin{activett}
	ham.getRate(511, {-413, -14, 13}, "Scheme2");
	ham.getRate("B0D*-MuNu", "Scheme2");
\end{activett}
both return the partial width for the $B^0 \to D^{*-}\mu^+\nu$ vertex, using the form factor parameterization specified in \textlst{"Scheme2"}, 
and using whatever WCs or FF uncertainties have been specified. 
For each WC specialization applied to weights via \textlst{applyWCSpecializationInWeights}, 
the corresponding rate tensor in the WC specialization basis is also computed. 
For an applied specialization \textlst{"JM"}, the corresponding rate can be retrieved by, e.g.
\begin{activett}
	ham.getRate("B0D*-MuNu", "Scheme2", "JM");
\end{activett}
using whatever WCs have been specified in that basis. 
The general rate is always computed by default, 
unless one changes the \textlst{"GeneralRates"} setting via \textlst{ham.setOptions("Hammer: {GeneralRates: false}");}

The \textlst{getRate} method is charge conjugate sensitive, so in a vertex string one must specify sufficient charges to make the vertex charge unique. 
(For example, writing just \textlst{"B0D*MuNu"} would have corresponded to not only $B^0 \to D^{*-}\mu^+\nu$, but also the (very heavily suppressed) process $B^0 \to D^{*+}\mu^-\bar{\nu}$.) 
The method \textlst{getDenominatorRate} takes just the vertex argument, 
and returns the partial width according to the specified denominator/input FF parametrization chosen in \textlst{setFFInputScheme},
and the denominator/input WCs. 
WC specializations are not possible for rates corresponding to the denominator/input FF schemes.

Vertices involving new physics and/or form factor parametrizations have rates implemented in dedicated classes, and integrated over $q^2$ (and other invariants as needed) via Gaussian quadrature. 
Other partial widths, e.g. for $D^* \to D\pi$ or $\tau \to \ell\nu\nu$, are obtained from the SM branching ratios and widths specified in the \textlst{Pdg} class. 
The partial width for each unique vertex is computed only \emph{once} per run, being computed and stored the first time each unique vertex is encountered in a process. 
Rates are computed vertex-wise inside edges. Hence, e.g., while an edge \paredge\, is computed as a single amplitude, the rates for the known and unknown vertices are computed and stored independently.
If a vertex is set to pure PS (or successfully set to pure PS inside an edge, see Sec.~\ref{sec:purePS}) then following the PS definition~\eqref{eqn:PSdef} the returned rate for that vertex is the phase space rate 
\begin{equation}
	\Gamma^{\mathcal{PS}}_n = \frac{1}{2m} \int m^{6-2n} d \mathcal{PS}_n\,.
\end{equation}

The rates for vertices whose amplitudes have no form factor---they are not required to be specified in a FF scheme---are 
automatically assigned to each scheme name relevant for the decay process to which they belong.
For example in a $B \to D (\tau \to \mu\nu\nu)\nu$ decay with schemes \textlst{"Scheme1"} and \textlst{"Scheme2"},
 the $\tau \to \mu \nu \nu$ partial width can be retrieved via \textlst{ham.getRate("Tau+Mu+NuNu", "Scheme1")} or \textlst{ham.getRate("Tau+Mu+NuNu", "Scheme2")}. 
Similarly, rates for vertices that have no WCs are automatically assigned to each WC specialization that has been applied.

\subsection{Multithreading}

The library has the ability to perform lock-free parallelization of the \textlst{getHistogram(s)} and \textlst{getWeight} evaluations.  
This requires the use of the thread-local methods \textlst{setWilsonCoefficientsLocal} and \textlst{setFFEigenvectorsLocal} to set the desired WC or FF uncertainties. 

The \textlst{...Local} methods take the same syntax as their global versions \textlst{setWilsonCoefficients} and \textlst{setFFEigenvectors}, but with different behaviour: 
They do not set the values incrementally from the current settings, but always increment from the SM and zero FF uncertainties, respectively. 
Global values of the WCs or FF variations are unaffected by the \textlst{...Local} methods, but the global \textlst{set...} methods should not be used in a multithreaded run.

\section{The Hammer buffer}
\label{sec:buf}
\hammer provides the ability to store header settings, generated event weights, histograms, and/or rates in binary buffers for later retrieval and reprocessing.
These buffers are built on the cross-platform serialization library \textlst{flatbuffers}: the binary buffer that needs to be read or written is wrapped in the struct \textlst{Hammer::IOBuffer}. 
The choice of how to implement the I/O is left to the user. To facilitate the use of streams, the operators \textlst{Hammer::IOBuffer::operator <<, >>} are available. 
To facilitate saving \hammer buffers into \textlst{TTree} leaves, one can use \textlst{Hammer::RootIOBuffer} as a wrapper around \textlst{Hammer::IOBuffer}.

\subsection{Saving}

As an example implementation of writing an \textlst{IOBuffer} into a file, we will use an \textlst{ofstream} 
\begin{activett}
	ofstream outFile("./DemoSave.dat",ios::binary);
\end{activett}

\subsubsection{Headers, events and rates}
\label{sec:saveheadevrate}
The methods \textlst{Hammer::save...} return a \textlst{IOBuffer}, which can be stored as a sequential record in the stream via an insertion (\textlst{<<}) operator. 
For example,
\begin{activett}
	outFile << ham.saveRunHeader();
\end{activett}
writes the declared run header, with all its settings, into an \textlst{IOBuffer} and passes it as a record into the buffer. 
 The available record types are labelled by an enum \textlst{char} \textlst{Hammer::RecordType} with values \textlst{UNDEFINED = 'u'}, \textlst{HEADER = 'b'}, \textlst{EVENT = 'e'}, \textlst{HISTOGRAM = 'h'}, 
 \textlst{HISTOGRAM_DEFINITION = 'd'}, and  \textlst{RATE = 'r'}. 
(Note the \textlst{saveOptionCard} method instead takes a filename and a bool for whether to write default values (\textlst{true}, default) versus modified settings (\textlst{false}). Output is written in text to the specified file.) Any combination of save methods may be invoked, in any order.

The method \textlst{saveEventWeights} saves the event weights of the currently initialized and processed event 
(there may be multiple weights saved if there are multiple processes in the event). 
This method should be invoked only after \textlst{processEvent}. Similarly, \textlst{saveRates} writes \emph{all} rates computed during the event loop,
including for all declared FF schemes and weight-applied WC specializations (see Sec.~\ref{sec:rates}).

\subsubsection{Histograms}
The method \textlst{saveHistogram}, when taking only a histogram name as an argument, saves the entire specified histogram set. The return value is the container \textlst{Hammer::IOBuffers}. An \textlst{operator <<} overload for \textlst{IOBuffers} allows to save all its entries at once (each histogram in a histogram set occupies an individual buffer record). For example,
\begin{activett}
	outFile << ham.saveHistogram("q2VsEmu");
\end{activett}
saves all the unique \textlst{"q2VsEmu"} histograms, corresponding to the unique event IDs and declared FF schemes, 
subject to compression settings (see Sec.~\ref{sec:compresshistos}).
Invocation of \textlst{saveHistogram} automatically also saves an additional separate buffer record for the histogram definition, immediately preceding the histogram record itself.
 
The \textlst{saveHistogram} method may optionally take additional arguments -- either an FF scheme name or an event ID group -- in order to save only part of an entire histogram set, 
if, e.g., space or file sizes are too large for an entire histogram set.
For instance, 
\begin{activett}
	outFile << ham.saveHistogram("q2VsEmu", "Scheme2");
\end{activett}
saves only those histograms in the histogram set computed for \textlst{Scheme2} and not for any other schemes that may have been declared. If event ID groups are known, one may instead save histograms for one group, \textlst{eventIDgroup}, via 
\begin{activett}
	outFile << ham.saveHistogram("q2VsEmu", eventIDgroup);
\end{activett}
Attempting to save a histogram that does not exist will result in an exception.

\subsubsection{ROOT}

Saving a buffer in \texttt{ROOT} format is achieved by passing the \textlst{IOBuffer} output of the \textlst{save...} methods into a  \textlst{RootIOBuffer}, which may then be stored in a \texttt{ROOT} \textlst{TTree}. 
Explicit implementations of this functionality are provided in various \textlst{demo...root.cc} example programs.

\subsection{(Re)loading}

Buffer records may be loaded from a declared \textlst{ifstream} infile into an \textlst{IOBuffer} via an \textlst{istream} operator, using \textlst{load...} methods 
with the same nomenclature as the \textlst{save...} methods. For example,
\begin{activett}
	ifstream inFile("./DemoSave.dat", ios::binary);
	Hammer::IOBuffer buf{Hammer::RecordType::UNDEFINED, 0ul, nullptr};
	inFile >> buf;
	ham.loadRunHeader(buf);
\end{activett}
attempts to load the first buffer record as a run header (returning \textlst{false} if this record is of a different type). 

It is the responsibility of the user to curate the logic and order under which a buffer is saved and then read. 
For example, if a block of histograms has been saved before a set of rate records, then
\begin{activett}
while(buf.kind != Hammer::RecordType::RATE) {
	if(buf.kind == Hammer::RecordType::HISTOGRAM) {
		ham.loadHistogram(buf);
	} 
	if(buf.kind == Hammer::RecordType::HISTOGRAM_DEFINITION){
		ham.loadHistogramDefinition(buf);
	}
	inFile >> buf;
}
\end{activett}
would read through the buffer, with the method \textlst{Hammer::loadHistogram} loading all the histograms, and \textlst{Hammer::loadHistogramDefinition} 
all the histogram definitions, that are found before reaching the block of saved rates. (One could instead have used 
\textlst{while(buf.kind \!= Hammer::RecordType::UNDEFINED)} to simply read through the entire buffer.)
The \textlst{loadHistogram} method in particular returns a struct \textlst{HistoInfo} with string variables for \textlst{name}, \textlst{scheme}, and \textlst{specialization}
and an event ID group (a set of event IDs, see Sec~\ref{sec:addhistos}) \textlst{eventGroupId}.

Once an object is loaded, it behaves just as the originally computed instance. 
Thus, one may invoke \textlst{getHistogram} for a reloaded histogram as described in Sec.~\ref{sec:retrievehistos}. 
(The method \textlst{removeHistogram} takes a histogram name and deletes that histogram from the instance.)

Event weights can be reloaded via \textlst{loadEventWeights}. This permits recreating the original event loop provided \textlst{initEvent} and \textlst{processEvent} are called appropriately. For example, on a block of saved event records
\begin{activett}
while(buf.kind == Hammer::RecordType::EVENT) {
	ham.initEvent();
	ham.loadEventWeights(buf);
	double q2 = ...; //Calculate q^2 from known kinematic event information 
	ham.fillEventHistogram("Q2", {q2});
	ham.processEvent();
	inFile >> buf;
}
\end{activett}			
would permit reprocessing of saved event weights into a newly created \textlst{"Q2"} histogram.

The method \textlst{loadRates} behaves similarly to \textlst{loadHistogram}. 
Event weights can be reloaded via \textlst{loadEventWeights}, recreating the original event loop provided \textlst{initEvent} and \textlst{processEvent} are called appropriately. For example,
 \begin{activett}
	inFile >> buf;
	while(buf.kind == Hammer::RecordType::EVENT) {
		ham.initEvent();
		ham.loadEventWeights(buf);
		double q2 = ...;
		ham.fillEventHistogram("Q2", {q2});
		ham.processEvent();
		inFile >> buf;
	}
\end{activett}			
would permit reprocessing of saved event weights into a newly created \textlst{"Q2"} histogram.

Loading a buffer in \texttt{ROOT} format is achieved by reading the \textlst{RootIOBuffer} stored in a \textlst{TTree} into an \textlst{IOBuffer} that can be passed to the \textlst{load...} methods. 
Explicit implementations of this functionality are provided in various \textlst{demo...root.cc} example programs.

\subsection{Parallelization and merging}
\label{sec:parallel_merge}

In order to permit parallelization of initialization analyses, the \textlst{load...} methods accept an additional bool, 
to specify whether to merge the buffer contents with existing objects in memory (\textlst{true}), or overwrite them (\textlst{false}, default). 

First, \textlst{loadRunHeader} permits merging of two sets of header settings into their union, with errors thrown for matching settings with non-matching values. 
When merging, if one wishes to be able to reprocess events (calculating weights etc),
the first invocation of \textlst{loadRunHeader} should be called with the merge bool set to \textlst{false}, and subsequent invocations set to \textlst{true}.
Otherwise, this and other load methods may be uniformly called with the merge bool \textlst{true}.

Merging of histograms occurs if two histograms are loaded with a matching name. 
This merging is \emph{additive} for histograms in each histogram set with the same event ID group and FF scheme, 
and otherwise results in the new unique histograms being \emph{appended} to the existing histogram set. (If one wishes instead to overwrite a histogram, one may instead first invoke
\textlst{removeHistogram}, and then reload the desired components of the histogram set.)

Errors are thrown if the matching histograms do not have compatible shapes or bin contents.
For instance, if the \textlst{"q2VsEmu"} histogram is loaded via
\begin{activett}
	inFile1 >> buf; 
	ham.loadHistogram(buf); 
\end{activett}	
subsequently loading an identically named histogram from a second infile via 
\begin{activett}
	inFile2 >> buf; 
	ham.loadHistogram(buf, true); 
\end{activett}
will merge the two histograms together according to the above rules.

The methods \textlst{loadEventWeights} and \textlst{loadRates} behave similarly. For weights (rates) with matching process ID (event ID), 
merging permits appending of process weights (rates) computed with new form factor schemes to the process weights (rates). 
In the case of merging rates, errors are thrown if a form factor scheme by the same name already exists for the same decay and the rate tensors do not match.

\acknowledgements
\hammer has been developed with the active participation of many users and colleagues.
The Hammer collaboration especially thanks for their extensive feedback, discussions, questions, and/or beta testing during development: 
Marco Colonna, Juli\'an Garc\'\i a Pardi\~nas, Lucia Grillo, Donal Hill, Emily Jiang, Anna Lupato, Abhijit Mathad, Simone Meloni, Biljana Mitreska, 
Adam Morris, Patrick Owen, Gary Robertson, Luke Scantlebury-Smead and Zishuo Yang from LHCb;
Henrik Junkerkalefeld, Kilian Lieret, Thomas Lueck, Tommy Martinov, Felix Metzner, Markus Prim, and Maximilian Welsch from Belle~II;
Riccardo Manzoni, Luigi Marchese, Patricia Wagner, and Yuta Takahashi from CMS.
We also thank Jeffrey Andersen for technical infrastructure support.
MP thanks the Aspen Center for Physics (supported by NSF Grant PHY-2210452) for its hospitality, where part of this work was performed. 
MP is supported by the Office of High Energy Physics of the U.S. Department of Energy under Award No. DE-SC0011632 and by the Walter Burke Institute for Theoretical Physics.
DJR is supported by the Office of High Energy Physics of the U.S. Department of Energy under contract DE-AC02-05CH11231.

\appendix
\makeatletter
\let\save@section\section
\renewcommand{\section}[1]{
	\save@section{#1}
	\addtocontents{toc}{\protect\vspace*{-10pt}}
}
\makeatother

\section{Conventions}
\label{sec:Conv}
\addtocontents{toc}{\protect\vspace*{10pt}}

\subsection{$V_{qb}$}
The $V_{cb}$ or $V_{ub}$ prefactor is generally not included explicitly in the $b \to c$ amplitudes, form factor parameters or rates. 
(One exception is the BGL parametrization, whose parameters typically absorb a factor of $V_{cb} \eta_{\text{EW}}$. 
In order to preserve uniformity among the form factor schemes, this factor is divided out of the BGL form factors.)

\subsection{NP operator basis}
\label{sec:NPops}
A complete basis for the four-Fermi operators mediating $b \to c \bar\ell \nu$ decay, including right-handed neutrinos, is shown in Table \ref{tab:NPc}. 
The NP couplings to the quark and lepton currents are denoted by $\tC^i_j$ and $\tL^i_j$, respectively, and may in general be complex numbers. 
The lower index of $\tL$ denotes the $\nu$ helicity and the lower index of $\tC$ is that of the $b$ quark. The NP couplings are normalized with respect to the SM current.
Note that the definitions below may be different from the library's internal WC basis conventions, which are required for the self-consistent addition of new amplitude classes extension of existing ones.

\begin{table}[h]
\renewcommand{\arraystretch}{1.25}
\newcolumntype{C}{ >{\centering\arraybackslash} m{2cm} <{}}
\def\simplecollect#1#2\ignorespaces#3\unskip{#1{#3}\unskip}
\newcommand*{\ctexttt}[1]{\centering{\textlst{#1}}}
\newcolumntype{S}{>{\simplecollect\ctexttt} m{3cm}}
\newcolumntype{D}{ >{\raggedright\arraybackslash $} c <{ \quad $}}
\begin{tabular}{C|SDD}
\hline
Current 				&  WC Tag 	& \text{WC}		& \text{4-Fermi}/ ( i2\sqrt{2}\, V_{cb} G_F)  \\
\hline\hline
 SM 					& SM		& 	1			&\big[\cbar \g^\mu P_L b\big] \big[\bar\ell \g_\mu P_L \nu\big] \\
 \hline
 \multirow{4}{*}{Vector} 	& V\_qLlL 		& \chVL \laVL		&\big[\cbar \chVL  \g^\mu P_L b\big] \big[\bar\ell \laVL \g_\mu P_L\nu\big] \\
 					& V\_qRlL		& \chVR \laVL		&\big[\cbar \chVR \g^\mu P_R b\big] \big[\bar\ell \laVL \g_\mu P_L\nu\big] \\
					& V\_qLlR		& \chVL \laVR	 	&\big[\cbar \chVL \g^\mu P_L b\big] \big[\bar\ell \laVR \g_\mu P_R\nu\big] \\
					& V\_qRlR		& \chVR \laVR		&\big[\cbar \chVR \g^\mu P_R b\big] \big[\bar\ell \laVR \g_\mu P_R\nu\big] \\
\hline
\multirow{4}{*}{Scalar} 	& S\_qLlL 		& \chSL\laSL 		&\big[\cbar \chSL P_L b\big] \big[\bar\ell \laSL P_L \nu\big] \\
					& S\_qRlL 	& \chSR\laSL		&\big[\cbar \chSR P_R b\big] \big[\bar\ell \laSL P_L \nu\big] \\
					& S\_qLlR 	& \chSL\laSR		&\big[\cbar \chSL P_L b\big] \big[\bar\ell \laSR P_R \nu\big] \\
					& S\_qRlR 	& \chSR\laSR		&\big[\cbar \chSR P_R b\big] \big[\bar\ell \laSR P_R \nu\big] \\
\hline
\multirow{2}{*}{Tensor} 	& T\_qLlL 		& \chTL\laTL		&\big[\cbar\, \chTL \sigma^\mn P_L b\big] \big[\bar\ell\, \laTL \sigma_\mn P_L \nu \big] \\
					& T\_qRlR 	& \chTR\laTR 		& \big[\cbar\, \chTR \sigma^\mn P_R b\big] \big[\bar\ell\, \laTR \sigma_\mn P_R \nu \big] \\
\hline
\end{tabular}
\caption{NP operator basis, and coupling conventions.}
\label{tab:NPc}
\end{table}

These conventions correspond to the conventions of Refs.~\cite{Ligeti:2016npd} via
\begin{align}
\chVL &= \alVL{}^*  \,, 		& \chVR & = \alVR{}^*\,, \nn\\*
\chSR &= -\alSL{}^*\,, 		& \chSL &= -\alSR{}^* \,, \nn\\*
\chTR &= -\alTL{}^*\,, 		& \chTL &= -\alTR{}^*\,, \nn\\*
\tL^{V,S,T}_L &= \beta^{V,S,T}_L{}^*\,,& \tL^{V,S,T}_R &= \beta^{V,S,T}_R{}^*\,.
\end{align}
All internal \hammer calculations are done in the $\alpha^i_j\beta^k_l$ basis of Ref.~\cite{Ligeti:2016npd}, which is naturally defined for $\bbar \to \cbar \ell \nu$ transitions and their corresponding $\bbar \Gamma c$ operators. Since, however, specification of WCs with respect to $\cbar \Gamma b$ operators is the predominant convention, \hammer inputs are specified in the $\tC^i_j\tL^k_l$ WC basis.
In the conventions of Ref.~\cite{Bernlochner:2017jxt}, $\chi = \tilde\alpha$, and $\lambda = \tilde\beta$, but we discard this tilded notation hereafter, so that there is no potential confusion as to which convention the WC tag subscripts, `\textlst{\_qXlX}', adhere.

\subsection{Lorentz signs}

For all amplitudes encoded into \hammer, we use a trace $-2$ metric, and the Lorentz sign conventions
\begin{equation}
	\text{Tr}[\g^\mu\g^\nu\g^\sigma\g^\rho\g^5] = -4i \epsilon^{\mu\nu\rho\sigma}\,,\qquad \epsilon^{0123} = +1\,.
\end{equation}
These choices fully specify all other possible ambiguous signs, for example the $\g^5$ trace choice is equivalent to $\sigma^\mn \g^5 \equiv +\frac{i}{2}\, \epsilon^{\mn \rho \sigma} \sigma_{\rho \sigma}$, with $\sigma_{\mn} = \frac{i}{2}[\g^\mu, \g^\nu]$\,.

\subsection{Form factor definitions}
\subsubsection{$\Bbar \to D$}
The $\Bbar \to D$ form factor tensor has ordered components
\begin{equation}
	\text{FF}_{D} = \Big\{f_S,\, f_0, \, f_+, \, f_T \Big\}\,,
\end{equation}
which are defined via
\begin{subequations}
\begin{align}
	 \ampBb{D}{\cbar\,b} & \equiv f_S \,, \\
	 \ampBb{D}{\cbar \g^\mu b} & \equiv f_+ (p_B + p_D)^\mu + [f_0 - f_+]\, \frac{m_B^2 - m_D^2}{q^2}\, q ^\mu\,, \\
	 \ampBb{D}{\cbar \sigma^\mn b} & \equiv i f_T \Big[(p_B + p_D)^{\mu}q^{\nu} - (p_B + p_D)^{\nu}q^{\mu}\Big]\,.
\end{align}
\end{subequations}
These definitions map to the conventional dimensionless form factor set $h_S, h_+, h_-, h_T$, as defined in e.g. Ref.~\cite{Bernlochner:2017jka}, via
\begin{subequations}
\label{eqn:BDhs}
\begin{align}
	f_S & = \sqrt{r_D} (w+1) m_B h_S\,, \\
	f_0 & = \frac{\sqrt{r_D}}{r_D^2-1}\big[ (r_D+1) (w-1) h_- +(r_D-1) (w+1) h_+ \big]\, \\
	f_+ & = \frac{(r_D-1) h_-+(r_D+1) h_+}{2\sqrt{r_D}}\,, \\
	f_T & = \frac{h_{\text{T}}}{2 \sqrt{r_D} m_B}\,,
\end{align}
\end{subequations}
with $r_D = m_D/m_B$. The $\Bbar \to D$ form factors $h_i$ are defined under the sign convention $\text{Tr}[\g^\mu\g^\nu\g^\sigma\g^\rho\g^5] = +4i \epsilon^{\mu\nu\rho\sigma}$, which is accounted for in eqs.~\eqref{eqn:BDhs}.

\subsubsection{$\Bbar \to D^*$}
The $\Bbar \to D^*$ form factor tensor has ordered components
\begin{equation}
	\text{FF}_{D^*} = \Big\{a_0,\, f,\, g,\, a_-,\, a_+,\, a_{T_0},\, a_{T_-},\, a_{T_+}  \Big\}\,,
\end{equation}
which are defined via
\begin{subequations}
\begin{align}
	 \ampBb{D^*}{\cbar \g^5 b} & \equiv a_0 \dotpr{\varepsilon^*}{p_B} \,,\\
	 \ampBb{D^*}{\cbar \g^\mu b} & \equiv -i g \, \epsilon^{\mn \rho \sigma}\, \varepsilon^*_\nu\, (p_B + p_{D^*})_\rho\, q_\sigma\,, \\
	 \ampBb{D^*}{\cbar \g^\mu \g^5 b} & \equiv  {\varepsilon^*}^\mu f + a_{+} \dotpr{\varepsilon^*}{p_B}(p_B + p_{D^*})^\mu + a_{-} \dotpr{\varepsilon^*}{p_B}q^\mu\,,  \\
	 \ampBb{D^*}{\cbar \sigma^\mn b} & \equiv -a_{T_+} \, \epsilon^{\mn \rho \sigma} \varepsilon^*_\rho (p_B + p_{D^*})_\sigma - a_{T_-}\, \epsilon^{\mn \rho \sigma} \varepsilon^*_\rho\, q_\sigma \notag \\*
	 & \quad - a_{T_0} \dotpr{\varepsilon^*}{p_B} \epsilon^{\mn \rho \sigma} (p_B + p_{D^*})_\rho\, q_\sigma\,.
\end{align}
\end{subequations}
These definitions map to the conventional dimensionless form factor set $h_P, h_V, h_{A_{1,2,3}}, h_{T_{1,2,3}}$, as defined in e.g. Ref.~\cite{Bernlochner:2017jka}, via
\begin{subequations}
\label{eqn:BDshs}
\begin{align}
	a_0 & = -\sqrt{r_{D^*}}h_{P}\,, \\
	f & = \sqrt{r_{D^*}} (w+1) m_B h_{A_1}\,, \\
	g & = \frac{h_{\text{V}}}{2 \sqrt{r_{D^*}} m_B}\,, \\
	a_- & = \frac{h_{A_3}-r_{D^*} h_{A_2}}{2 \sqrt{r_{D^*}} m_B}\,, \\
	a_+ & = -\frac{r_{D^*} h_{A_2}+h_{A_3}}{2 \sqrt{r_{D^*}} m_B}\,, \\
	a_{T_0} & = \frac{h_{T_3}}{2 \sqrt{r_{D^*}} m_B^2}\,, \\
	a_{T_-} & = \frac{(1-r_{D^*}) h_{T_1}-(r_{D^*}+1) h_{T_2}}{2 \sqrt{r_{D^*}}}\,, \\
	a_{T_+} & = \frac{(1 - r_{D^*}) h_{T_2} - (r_{D^*}+1) h_{T_1}}{2\sqrt{r_{D^*}}}\,.
\end{align}
\end{subequations}
with $r_{D^*} = m_{D^*}/m_B$. The $\Bbar \to D^*$ form factors $h_i$ are defined under the sign convention $\text{Tr}[\g^\mu\g^\nu\g^\sigma\g^\rho\g^5] = +4i \epsilon^{\mu\nu\rho\sigma}$, which is accounted for in eqs.~\eqref{eqn:BDshs}.

\subsubsection{$B \to D^{**}$}
The $B \to \dss$ form factor tensors are ordered
\begin{subequations}
\begin{align}
	\text{FF}_{\dSs} & = \Big\{g_P,\, g_+, \, g_-, \, g_T \Big\}\,, \\
	\text{FF}_{\dVs} & = \Big\{g_S,\, g_{V_1},\, g_{V_2},\, g_{V_3},\, g_a,\, g_{T_1},\, g_{T_2},\, g_{T_3} \Big\}\,,\\
	\text{FF}_{\dVp} & = \Big\{f_S,\, f_{V_1},\, f_{V_2},\, f_{V_3},\, f_a,\, f_{T_1},\, f_{T_2},\, f_{T_3} \Big\}\,,\\
	\text{FF}_{\dTs} & = \Big\{k_P,\, k_{A_1},\, k_{A_2},\, k_{A_3},\, k_V,\, k_{T_1},\, k_{T_2},\, k_{T_3} \Big\}\,,
\end{align}
\end{subequations}
which following Ref.~\cite{Bernlochner:2017jxt}, are defined for $\Bbar \to \dSs$ via
\begin{subequations}
\begin{align}
\ampBb{\dSs}{\cbar\,b} &= \ampB{\dSs}{\cbar \g_\mu b} = 0\,, \nn\\
\ampBb{\dSs}{\cbar\g_5 b} &= \sqrt{m_{\dSs} m_B}\, g_P\,, \nn\\
\ampBb{\dSs}{\cbar \g_\mu \g_5 b} &= \sqrt{m_{\dSs} m_B} 
  \big[ g_+ (v_\mu+v'_\mu) + g_- (v_\mu-v'_\mu) \big] , \nn\\
\ampBb{\dSs}{\cbar \sigma_{\mu\nu} b} &= 
  \sqrt{m_{\dSs} m_B}\, g_T\, \varepsilon_{\mu\nu\alpha\beta}\, 
  v^\alpha v'^\beta \,,
\end{align}
for $\Bbar \to \dVs$,
\begin{align}
\ampBb{\dVs}{\cbar\,b} &= -\sqrt{m_{\dVs} m_B}\, g_S\, (\epsilon^*\cdot v)\,, \nn\\
\ampBb{\dVs}{\cbar\g_5 b} &= 0\,, \nn\\
\ampBb{\dVs}{\cbar \g_\mu b} &=  \sqrt{m_{\dVs} m_B}\, 
  \big[ g_{V_1}\, \epsilon^*_\mu + (g_{V_2} v_\mu + g_{V_3} v'_\mu)\, (\epsilon^* \cdot v) \big]\,, \nn\\
\ampBb{\dVs}{\cbar \g_\mu \g_5 b} &= i \sqrt{m_{\dVs} m_B}\, g_A\, 
  \varepsilon_{\mu\alpha\beta\gamma}\, \epsilon^{*\alpha} v^\beta\, v'^\gamma\,,\\
\ampBb{\dVs}{\cbar \sigma_{\mu\nu} b} &= i \sqrt{m_{\dVs} m_B}\,
  \big[ g_{T_1} (\epsilon^*_\mu v_\nu - \epsilon^*_\nu v_\mu) + g_{T_2} (\epsilon^*_\mu v'_\nu - \epsilon^*_\nu v'_\mu) + g_{T_3} (\epsilon^*\!\cdot\! v) (v_\mu v'_\nu - v_\nu v'_\mu) \big]. \nn
\end{align}
for $\Bbar \to \dVp$,
\begin{align}\label{formf321}
\ampBb{\dVp}{\cbar\,b} &= \sqrt{m_{\dVp} m_B}\, f_S\, (\epsilon^*\cdot v) \,,\nn\\
\ampBb{\dVp}{\cbar\g_5 b} &= 0\,, \nn\\
\ampBb{\dVp}{\cbar\g_\mu b} &= \sqrt{m_{\dVp} m_B}\, \big[ f_{V_1}\, \epsilon^*_\mu + (f_{V_2} v_\mu + f_{V_3} v'_\mu) (\epsilon^*\cdot v) \big] , \nn\\
\ampBb{\dVp}{\cbar\g_\mu\g_5 b} &= i\, \sqrt{m_{\dVp} m_B}\, f_A\,
\varepsilon_{\mu\alpha\beta\gamma} \epsilon^{*\alpha} v^\beta v'^\gamma \,,\\
\ampB{\dVp}{\cbar \sigma_{\mu\nu} b} &= i \sqrt{m_{\dVp} m_B} \big[ f_{T_1}
(\epsilon^*_\mu v_\nu - \epsilon^*_\nu v_\mu) + f_{T_2} (\epsilon^*_\mu v'_\nu - \epsilon^*_\nu v'_\mu) + f_{T_3} (\epsilon^*\!\cdot \! v) (v_\mu v'_\nu - v_\nu v'_\mu) \big], \nn
\end{align}
and finally for  $\Bbar \to \dTs$,
\begin{align}
\ampBb{\dTs}{\cbar\,b} &= 0 \,, \nn\\
\ampBb{\dTs}{\cbar\g_5 b} &= \sqrt{m_{\dTs} m_B}\,
  k_P\,\epsilon^*_{\alpha\beta}\, v^\alpha v^\beta \,, \nn\\
\ampBb{\dTs}{\cbar\g_\mu b} &= i \sqrt{m_{\dTs} m_B}\, k_V\,
  \varepsilon_{\mu\alpha\beta\gamma}\,\epsilon^{*\alpha\sigma} v_\sigma v^\beta
  v'^\gamma \,,\nn\\
\ampBb{\dTs}{\cbar\g_\mu\g_5 b}  &= \sqrt{m_{\dTs} m_B}\, \big[ k_{A_1}\,
  \epsilon^*_{\mu\alpha} v^\alpha + (k_{A_2} v_\mu + k_{A_3} v'_\mu)\,
  \epsilon^*_{\alpha\beta}\, v^\alpha v^\beta \big] , \\
\ampBb{\dTs}{\cbar \sigma_{\mu\nu} b}  &= \sqrt{m_{\dTs} m_B}\,
  \varepsilon_{\mu\nu\alpha\beta}\, \big\{ [k_{T_1} (v+v')^\alpha + k_{T_2} (v-v')^\alpha)]\, \epsilon^{*\gamma\beta} v_\gamma + k_{T_3}\, v^\alpha v'^\beta \epsilon^{*\rho\sigma} v_\rho v_\sigma\big\}. \nn
\end{align}
\end{subequations}

(NB: In the case of the \textlst{ISGW2} FF parametrization for the $\dVp$ and $\dVs$, \texttt{EvtGen} includes an additional \emph{ad hoc} `smearing' by the factor 
$\sqrt{q^2_{\text{max}, \text{mean}}/q^2_{\text{max}}}$ on each form factor.
This is included by default, but can be deactivated via the bool setting \textlst{"SmearQ2"}.)

\subsubsection{$\Lb \to \Lc$}
The $\Lb^0 \to \Lc^+$ form factor tensor has ordered components
\begin{equation}
	\text{FF}_{\Lc} = \Big\{ h_{S},\,h_{P},\,f_{1},\,f_{2},\,f_{3},\,g_{1},\,g_{2},\,g_{3},\,h_{1},\,h_{2},\,h_{3},\,h_{4}\Big\}\,,
\end{equation}
The form factors are defined as in Ref.~\cite{Bernlochner:2018bfn}, using the sign convention $\text{Tr}[\g^\mu\g^\nu\g^\sigma\g^\rho\g^5] = -4i \epsilon^{\mu\nu\rho\sigma}$, via
\begin{subequations}
\label{LcHQETffdef} 
\begin{align}
\langle \Lambda_c(p',s')| \bar c\, b |\Lambda_b(p,s)\rangle
  &= h_S\, \bar u(p',s')\, u(p,s)\,, \\
\langle \Lambda_c(p',s')| \bar c \gamma_5 b |\Lambda_b(p,s)\rangle
  &= h_P\, \bar u(p',s')\, \gamma_5\, u(p,s)\,, \\
\langle \Lambda_c(p',s')| \bar c\gamma_\nu b |\Lambda_b(p,s)\rangle
  &= \bar u(p',s') \big[ f_1 \gamma_\mu + f_2 v_\mu + f_3 v'_\mu \big]
  u(p,s)\,, \\
\langle \Lambda_c(p',s')| \bar c\gamma_\nu\gamma_5 b |\Lambda_b(p,s)\rangle
  &= \bar u(p',s') \big[ g_1 \gamma_\mu + g_2 v_\mu + g_3 v'_\mu \big] 
  \gamma_5\, u(p,s)\,,\\
\langle \Lambda_c(p',s')| \bar c\, \sigma_{\mu\nu}\, b |\Lambda_b(p,s)\rangle
  &= \bar u(p',s') \big[ h_1\, \sigma_{\mu\nu}
  + i\, h_2 (v_\mu \gamma_\nu - v_\nu \gamma_\mu)
  + i\, h_3 (v'_\mu \gamma_\nu - v'_\nu \gamma_\mu) \\
  & \qquad\quad + i\, h_4 (v_\mu v'_\nu - v_\nu v'_\mu) \big] u(p,s)\,.
\end{align}
\end{subequations}
The spinors are normalized to $\bar u(p,s)u(p,s) = 2m$. (Note that another common definition for the SM form factors is~\cite{Falk:1992ws}
\begin{align}
\label{LcQCDffdef}
\langle \Lambda_c(p',s')| \bar c\gamma_\mu b |\Lambda_b(p,s)\rangle
  &= \bar u(p',s') \big[ F_1\, \gamma_\mu - i F_2\, \sigma_{\mu\nu}\, q^\nu
  + F_3\, q_\mu \big] u(p,s)\,, \nn\\
\langle \Lambda_c(p',s')| \bar c\gamma_\mu\gamma_5 b |\Lambda_b(p,s)\rangle 
  &= \bar u(p',s') \big[ G_1\, \gamma_\mu - i G_2\, \sigma_{\mu\nu}\, q^\nu
  + G_3\, q_\mu \big] \gamma_5\, u(p,s)\,.
\end{align}
The notation of Ref.~\cite{Falk:1992ws} also exchanges upper and lowercase symbols -- i.e. $F_i\leftrightarrow f_i$ and $G_i\leftrightarrow g_i$ -- with respect to Eqs.~\eqref{LcHQETffdef} and~\eqref{LcQCDffdef}.)

\subsubsection{$B \to \rho$, $\omega$}
The $B \to \rho$ or $\omega$ decays have form factor tensor with ordered components
\begin{equation}
	\text{FF}_{\rho,\omega} = \Big\{ A_P, V, A_0, A_1, A_{12}, T_1, T_2, T_{23}\Big\}\,,
\end{equation}
which are defined as in Ref.~\cite{Straub:2015ica}. Explicitly,
\begin{subequations}
\begin{align}
	 \sqrt{2}\ampBb{V}{\ubar \g^5 b} & \equiv A_P \dotpr{\varepsilon^*}{q} \,,\\
	\sqrt{2} \ampBb{V}{\ubar \g^\mu b} & \equiv - \frac{i V}{m_B + m_V} \, \epsilon^{\mn \rho \sigma}\, \varepsilon^*_\nu\, (p_B + p_V)_\rho\, q_\sigma\,, \\
	\sqrt{2} \ampBb{V}{\ubar \g^\mu \g^5 b} & \equiv  A_1(m_B + m_V) \varepsilon^{*\mu} - A_2\frac{(p_B + p_V)^\mu \dotpr{\varepsilon^*}{q}}{m_B + m_V}  \\
	& \quad + \frac{\dotpr{\varepsilon^*}{q} q^\mu}{q^2}\Big[ A_2(m_B - m_V) - A_1(m_B + m_V) + 2 m_V A_0 \Big] \,,\\
	\sqrt{2} \ampBb{V}{\ubar \sigma^\mn b} & \equiv  \epsilon^{\mn \rho \sigma}\bigg[ T_1 \varepsilon^*_\rho (p_B + p_V)_\sigma - (T_2 + T_1)\frac{m_B^2 - m_V^2}{q^2}\varepsilon^*_\rho q_\sigma \\
	& \qquad + (p_B + p_V)_\rho q_\sigma \frac{\dotpr{\varepsilon^*}{q}}{q^2} \bigg((T_1 + T_2)  + T_3 \frac{q^2}{m_B^2 - m_V^2}\bigg)\bigg] \,,
\end{align}
\end{subequations}
with the additional redefinitions with respect to $A_{12}$ and $T_{23}$,
\begin{subequations}
\begin{align}
	A_2 & = \frac{A_1(m_B^2 - m_V^2 -q^2)(m_B + m_V)^2 - 16 A_{12} m_B m_V^2(m_B + m_V)}{4|p_V|^2 m_B^2}\,,\\
	T_3 & = \frac{T_2(m_B^2 + 3m_V^2 -q^2)(m_B^2 - m_V^2) - 8 T_{23} m_B m_V^2(m_B - m_V)}{4|p_V|^2 m_B^2}\,,
\end{align}
\end{subequations}
with $|p_V|$ the vector meson 3-momentum in the $B$ rest frame. 

Light-cone sum rule results (LCSR) results are available for both the SM and NP form factors, parametrized by an optimized $z$ expansion, 
in the form of a correlated, beyond zero recoil fit between the SM and NP form factors~\cite{Straub:2015ica}. 
This LCSR-based parametrization is referred to as `\textlst{BSZ}'. 

\subsubsection{$\Lb \to \Lc^*(2595)$ and $\Lc^*(2625)$}
\newcommand{\dS}{d_{S}}
\newcommand{\dP}{d_{P}}
\newcommand{\dV}[1]{d_{V#1}}
\newcommand{\dA}[1]{d_{A#1}}
\newcommand{\dT}[1]{d_{T#1}}
\newcommand{\lS}{l_{S}}
\newcommand{\lP}{l_{P}}
\newcommand{\lV}[1]{l_{V#1}}
\newcommand{\lA}[1]{l_{A#1}}
\newcommand{\lT}[1]{l_{T#1}}
\newcommand{\LcSp}{
	\mathchoice{\Lambda_c^*(2595)}
			{\Lambda_c^*(2595)}{\Lambda_c^*(2595)}{\Lambda_c^*(2595)}
}
\newcommand{\LcTn}{
	\mathchoice{\Lambda_c^*(2625)}
		{\Lambda_c^*(2625)}{\Lambda_c^*(2625)}{\Lambda_c^*(2625)}
}
\newcommand{\Lcs}{\Lambda_c^*}
\newcommand{\Psic}{{\Psi\kern-0.1em}_c}
\newcommand{\Psicbar}{{\overline{\phantom{h}}\kern-0.6em\Psi\kern-0.1em}_c}
\newcommand{\ccdot}{\!\cdot\!}
The $\Lb^0 \to \Lc^*$ form factor tensors have ordered components
\begin{subequations}
\begin{align}
	\text{FF}_{\LcSp} & = \Big\{ \dS, \dP, \dV1, \dV2, \dV3, \dA1, \dA2, \dA3, \dT1, \dT2, \dT3, \dT4 \Big\}\,, \\
	\text{FF}_{\LcTn} & = \Big\{ \lS, \lP, \lV1, \lV2, \lV3, \lV4, \lA1, \lA2, \lA3, \lA4, \lT1, \lT2, \lT3, \lT4, \lT6, \lT7 \Big\}\,.
\end{align}
\end{subequations}
following the conventions and definitions in Ref.~\cite{Papucci:2021pmj}. 
The form factor $\lT5$ must be eliminated, 
according to the kernel of the $\Lb \to \Lc^*(2625)$ amplitudes, after matching onto HQET or a particular model of interest. 
See Ref.~\cite{Papucci:2021pmj}.

Explicitly, representing $\Lb$ and $\LcSp$ by spinors $u_b(p,s)$ and $\bar u_c(p',s')$, respectively, with momenta  $p = m_{\Lb}v$ and $p' = m_{\Lc^*}v'$,
the form factors $d_X$ are defined via
\begin{align}
	\langle \LcSp | \bar c\, b |\Lb\rangle &= -\dS\, \bar u_c \g_5 u_b\,, \\
	\langle \LcSp | \bar c \g_5 b |\Lb \rangle &= -\dP \, \bar u_c\, u_b\,, \nn\\*
	\langle \LcSp | \bar c\g_\mu b |\Lb \rangle
		&= \bar u_c \big[ \dV1 \g_\mu + \dV2 v_\mu + \dV3 v'_\mu \big]\g_5 u_b\,, \nn\\
	\langle \LcSp | \bar c\g_\mu\g_5 b |\Lb \rangle
 		 &= \bar u_c \big[ \dA1 \g_\mu + \dA2 v_\mu + \dA3 v'_\mu \big] u_b\,, \nn \\
	\langle \LcSp | \bar c\, \sigma_{\mn}\, b |\Lb \rangle
  		&= -\bar u_c \big[ \dT1\, \sigma_{\mn} + i\, \dT2 v_{[\mu} \g_{\nu]}  + i\, \dT3 v'_{[\mu} \g^{\phantom{\prime}}_{\nu]} + i\, \dT4 v^{\phantom{\prime}}_{[\mu} v'_{\nu]} \big] \g_5 u_b\,, \nn 	 
\end{align}
The charmed spin-$3/2$ state is represented by a Rarita-Schwinger tensor, $\Psic^\mu(p',s')$, 
satisfying the usual transversity and projective conditions $v' \ccdot \Psic = 0$ and $\gamma \ccdot \Psic = 0$. 
The form factors $l_X$ are then defined via
\begin{align}
	\langle \LcTn | \bar c\, b |\Lb\rangle &= \lS\, v \ccdot \Psicbar  u_b\,, \\
	\langle \LcTn | \bar c \g_5 b |\Lb \rangle &= \lP \, v\ccdot \Psicbar \g_5 u_b\,, \nn\\*
	\langle \LcTn | \bar c\g_\mu b |\Lb \rangle
		&= v\ccdot \Psicbar \big[ \lV1 \g_\mu + \lV2 v_\mu + \lV3 v'_\mu \big] u_b + \lV4 {\Psicbar}_\mu u_b \,, \nn \\
	\langle \LcTn | \bar c\g_\mu\g_5 b |\Lb \rangle
 		 &= v\ccdot \Psicbar \big[ \lA1 \g_\mu + \lA2 v_\mu + \lA3 v'_\mu \big] \g_5 u_b + \lA4 {\Psicbar}_\mu \g_5 u_b \,, \nn \\
	\langle \LcTn | \bar c\, \sigma_{\mn}\, b |\Lb \rangle
  		&= v\ccdot \Psicbar \big[ \lT1\, \sigma_{\mn} + i\, \lT2 v_{[\mu} \g_{\nu]} + i\, \lT3 v'_{[\mu} \g^{\phantom{\prime}}_{\nu]} + i\, \lT7 v^{\phantom{\prime}}_{[\mu} v'_{\nu]} \big]u_b  \nn\\
		&+ i\, {\Psicbar}_{[\mu} \big[ \lT4\, \g^{\phantom{\prime}}_{\nu]} + \lT5 v^{\phantom{\prime}}_{\nu]} + \lT6 v'_{\nu]} \big] u_b \,. \nn
\end{align}

\subsection{Form factor variations}
\label{sec:FFUC}
At present, e.g. the \textlst{BGLVar} parametrization permits via \textlst{setFFEigenvectors} functionality (see Sec.~\ref{sec:WCFF})
direct manipulation of the $a_i$, $b_i$, $c_i$ and $d_i$ parameters for $B \to D^*$ 
(also denoted $a^g_i$, $a^f_i$, $a^{\mathcal{F}_1}_i$ and $a^{\mathcal{P}_1}_i$, respectively, in some notational conventions).
That is, the covariance matrix is set to the identity, and the basis of variations
\begin{activett}
	{"delta_a0","delta_a1","delta_a2","delta_b0","delta_b1","delta_b2",
				"delta_c1","delta_c2", "delta_d0","delta_d1"} ,
\end{activett}
Similarly the $a^{f_+}_i$ and $a^{f_0}_i$, $i =0,\ldots,3$ parameters are directly manipulated for $B \to D$ (also denoted $a_i$ and $b_i$, respectively, in some notational conventions), 
with respect to the basis
\begin{activett}
	{"delta_ap0","delta_ap1","delta_ap2","delta_ap3",
				"delta_a00","delta_a01","delta_a02","delta_a03"} .
\end{activett}

By contrast, the linearized form of the $B \to \rho$, $\omega$ LCSR-based parametrization, referred to as `\textlst{BSZVar}', features a non-trivial covariance matrix. 
At present we include just the first eight principal directions of the $21$ parameter fit, in the basis
\begin{activett}
	{"delta_e1","delta_e2","delta_e3","delta_e4","delta_e5","delta_e6",
				"delta_e7","delta_e8"} .
\end{activett}
Each covariance eigenvector $e_i$ is normalized by the square-root of its eigenvalue $\sqrt{\lambda_i}$, so that unit variation in each \textlst{"delta_ei"} corresponds to a $1\sigma$ variation.

A large number of other \textlst{Var} classes are available in the library: the implemented basis of parameters may be easily read off each class implementation.

\subsection{$D^{**}$ strong decays}

The library incorporates the strong decays $D_1 \to D^* \pi$, $D_2^* \to D^{(*)}\pi$. While the latter can proceed only via $d$-wave, 
the former proceeds by $d$-wave at leading order in HQET but may include $s$-wave contributions at subleading order~\cite{Casalbuoni:1996pg,Kilian:1992hq,Lu:1991px}, 
that may be thought of as a contribution arising from $D_1^*$--$D_1$ mixing. 
When $D^{**}$ decays are included in a run, a `form-factor' parametrization for them must be specified in each FF scheme: At present, the only partial-wave parametrization `\textlst{PW}' is included. 

The explicit $D_1 \to D^* \pi$ partial-wave amplitude
\begin{align}
	\mathcal{A}^{\lambda\kappa}_{D_1^+ \to D^* \pi} 
	 & = \frac{1}{f_\pi} \bigg(\frac{D}{\sqrt{6}} + \sqrt{\frac{3}{2}} S \bigg)\bigg[\frac{E_{D^*}^* - m_{D^*}}{m_{D_1}} \varepsilon^{-\lambda}_1 \ccdot p_{\pi} \ \varepsilon_*^{\kappa} \ccdot p_{\pi} 
	 	+ |\bm{p}_{\pi}|^2 \varepsilon_1^{-\lambda} \ccdot\varepsilon^\kappa_*\bigg]\\
		& \qquad + \frac{D}{\sqrt{6}} \frac{3m_{D^*}}{m_{D_1}}\varepsilon^{-\lambda}_1 \ccdot p_{\pi} \ \varepsilon_*^{\kappa} \ccdot p_{\pi} \,,
\end{align}
in which $|\bm{p}_{\pi}|$ is the pion momentum in the $D_1$ frame, and the $S$ and $D \in \mathbb{C}$ parametrize the $s$-~and $d$-wave contributions, respectively.
This is equivalent to the conventions in Ref.~\cite{Lu:1991px}, but with $S$ scaled by an additional factor $-3/\sqrt{2}$ with respect to $D$ 
to match the conventions of the \texttt{EvtGen} `\texttt{VVSPWave}' class. ($S$ is normalized by an additional $|\bm{p}_{\pi}|$ factor so that it is dimensionless.)
The corresponding partial rate $\Gamma_{D_1 \to D^* \pi}  = (|D|^2 + 9|S|^2/2) |\bm{p}_{\pi}|^5/(24\pi f_{\pi}^2 m_{D_1}^2)$.
The parameters $S$ and $D$ are treated as constant form-factors by the library, and may be set as options of the `\textlst{PW}' parametrization. 
This permits reweighting between specific $d$/$s$-wave admixtures. The default is $S=0$, $D=1$.
 
The explicit $D_2^* \to D^{(*)} \pi$ partial-wave amplitudes, in the same notation are
\begin{align}
	\mathcal{A}^{\lambda\kappa}_{D_2^{*+} \to D^* \pi}  & = \frac{D}{f_\pi} i\epsilon^{\alpha\beta\gamma\delta} 
		\varepsilon_{\alpha\tau}^{-\lambda}p_\pi^\tau  {\epsilon^\kappa_{D^*}}_\beta {p_{\pi}}_\gamma {p_{D^*}}_\delta\,,\\
	\mathcal{A}^{\lambda}_{D_2^{*+} \to D \pi}  & = \frac{D}{f_\pi} \varepsilon_{\mu\nu}^{-\lambda} p_\pi^\mu p_\pi^\nu\,,
\end{align}
corresponding to the partial rates $\Gamma_{D_2^* \to D^* \pi}  = |D|^2 |\bm{p}_{\pi}|^5/(40\pi f_{\pi}^2 m_{D_1}^2)$ 
and $\Gamma_{D_2^* \to D \pi}  = |D|^2 |\bm{p}_{\pi}|^5/(60\pi f_{\pi}^2 m_{D_1}^2)$, respectively.
The parameter $D$ is treated as a constant form factor, and may be set as an option of the `\textlst{PW}' parametrization. The default is $D=1$.

\subsection{Resonance lineshapes}
\texttt{EvtGen} includes additional momentum and angular momentum dependent models for resonance lineshapes, which can be numerically non-negligible for broad resonances such as the $D^{**}$. 
These (somewhat arbitrary) lineshape models typically factorize from the decay amplitudes themselves, and are generically invariant under reweighting.
They are therefore not presently included automatically the \hammer library: 
Effects of reweighting on the lineshape model, if important, can instead be included by the user via \textlst{setEventBaseWeight}.
The latter may be required in two cases:
\begin{itemize}
	\item \texttt{EvtGen} does not incorporate the lineshape in the case the decay is pure phase space. 
	Thus, caution should be used when reweighting broad resonances from pure phase generated by \texttt{EvtGen}, because lineshape weights included by \texttt{EvtGen} will be absent.
	\item The \texttt{EvtGen} lineshape models are sensitive to the angular momentum of the resonance. 
	Thus, reweighting that alters an admixture of partial waves (such as is possible with the \textlst{PW} FF parametrization of the $D_1 \to D^*\pi$ decay) may incorporate a different lineshape than would have been generated directly by \texttt{EvtGen}.
\end{itemize}
For more information on lineshape options in \texttt{EvtGen} we refer to its documentation~\cite{Ryd:2005zz}.




\section{Installation}

Detailed installation instructions are provided in the \textlst{README.md} accompanying the source code, 
and can be found on the \href{https://hammer.physics.lbl.gov}{webpage} or \href{https://gitlab.com/mpapucci/Hammer}{git repository}.

\section{API methods, operators, and structures}
A summary of API methods or operators belonging to the user-facing classes \textlst{Hammer}, 
\textlst{Process}, \textlst{Particle}, \textlst{FourMomentum}, \textlst{IOBuffer}, \textlst{IOBuffers} and \textlst{Log} are shown in Table~\ref{tab:api}.
The user-facing \textlst{enum} classes \textlst{WTerm} and \textlst{PAction} take values as
\begin{activett}
	WTerm :	{NUMERATOR, DENOMINATOR, COMMON}
	PAction :	{ALL, WEIGHTS, HISTOGRAMS}
\end{activett}
The \textlst{RecordType} \textlst{enum} of \hammer buffer types (the \textlst{kind} member of the \textlst{IOBuffer} class) takes \textlst{char} values
\begin{activett}
    RecordType :	{UNDEFINED = 'u', HEADER = 'b', EVENT = 'e', HISTOGRAM = 'h', RATE = 'r', HISTOGRAM_DEFINITION = 'd'}
\end{activett}
In addition, the user-facing struct \textlst{BinContents} has variables
\begin{activett}
	double sumWi  // sum of weights
	double sumWi2 // sum of squared weights
	size_t n      // number of entries in the bin
\end{activett}
and the struct \textlst{HistoInfo} has variables
\begin{activett}
	string name
	string scheme
	string specialization
	set<set<size_t>> eventGroupId 	// set of event IDs contributing to histogram
\end{activett}

\begin{table}[htb]
\begin{center}
\def\simplecollect#1#2\ignorespaces#3\unskip{#1{#3}\unskip}
\newcommand*{\textlsttab}[1]{\raggedright{\textlst{#1}}}
\newcolumntype{S}{>{\simplecollect\textlsttab} m{6cm}}
\newcolumntype{C}{ >{\centering\arraybackslash} m{1cm} <{}}
\newcommand{\Nskip}[1]{{\CPPidentifierstyle #1}}
\renewcommand{\arraystretch}{0.8}
\scalebox{0.75}{\parbox{1.33\linewidth}{
\begin{tabular}[t]{|SC|}
\hline
\multicolumn{1}{|c}{\textlst{Hammer} class API methods} & \\
\hline\hline
addFFScheme & \\
addHistogram & \\
addProcess & \\
addPurePSVertices & \\
addTotalSumOfWeights & \\
appliedWCSpecializationsInHistogram & \\
appliedWCSpecializationsInWeights & \\
applyWCSpecializationInAllHistograms & \\
applyWCSpecializationInHistogram & \\
applyWCSpecializationInWeights & \\
availableWCSpecializationIds & \\
clearPurePSVertices & \\
collapseProcessesInHistogram & \\
createProjectedHistogram & \\
createWCSpecialization & \\
fillEventHistogram & \\
forbidDecay & \\
getDenominatorRate & \\
getFFSchemeNames & \\
getHistogram & \\
getHistogram1D|!{\CPPidentifierstyle{2D}}\lstinline! |!{\CPPidentifierstyle{3D}}\lstinline!  & \\
getHistogramBinEdges & \\
getHistogramEventIds & \\
getHistogramShape & \\
getHistograms & \\
getHistograms1D|!{\CPPidentifierstyle{2D}}\lstinline! |!{\CPPidentifierstyle{3D}}\lstinline! & \\
getRate & \\
getWeight, getWeights & \\
histogramHasUnderOverFlows & \\
includeDecay & \\
initEvent & \\
initRun & \\
keepErrorsInHistogram & \\
loadEventWeights & \\
loadHistogram & \\
loadHistogramDefinition & \\
loadRates & \\
loadRunHeader & \\
processEvent & \\
readCards & \\
reconcileGeneralHistogramSetting & \\
reconcileGeneralWeightSetting & \\
reconcileSpecializations & \\
removeAllWCSpecializations & \\
removeFFScheme & \\
removeFFSpecializationInHistogram & \\
\hline
\end{tabular}%
\begin{tabular}[t]{|SC|}
\hline
\multicolumn{1}{|c}{\textlst{Hammer} class API methods} & \\
\hline\hline
removeHistogram & \\
removeProcess & \\
removeWCSpecialization & \\
removeWCSpecializationInHistogram & \\
removeWCSpecializationInWeights & \\
renameFFEigenvectors & \\
resetFFEigenvectors & \\
resetWilsonCoefficients & \\
retrieveFFEigenvectors & \\
retrieveWilsonCoefficients & \\
saveEventGeneralWeights & \\
saveEventWeights & \\
saveHeaderCard & \\
saveHistogram & \\
saveOptionCard & \\
saveRates & \\
saveReferences & \\
saveRunHeader & \\
setEventBaseWeight & \\
setEventHistogramBin & \\
setFFEigenvectors & \\
setFFEigenvectorsLocal & \\
setFFInputScheme & \\
setHeader & \\
setHistogram1D|!{\CPPidentifierstyle{2D}}\lstinline! |!{\CPPidentifierstyle{3D}}\lstinline! & \\
setHistograms1D|!{\CPPidentifierstyle{2D}}\lstinline! |!{\CPPidentifierstyle{3D}}\lstinline! & \\
setOptions & \\
setUnits & \\
setWCSpecializationBasis & \\
setWCSpecializationCoord & \\
setWCSpecializationOrigin & \\
setWilsonCoefficients & \\
setWilsonCoefficientsLocal & \\
showAvailableFFParams & \\
specializeFFInHistogram & \\
\hline
\multicolumn{1}{|c}{\textlst{Process} class API methods} & \\
\hline\hline
addParticle & \\
addVertex & \\
getParticlesByVertex & \\
getVertexId & \\
removeVertex & \\
\hline
\multicolumn{1}{|c}{\textlst{Particle} class API methods} & \\
\hline\hline
fromRoot & \\
momentum, p & \\
pdgId & \\
\hline
\end{tabular}%
\begin{tabular}[t]{|SC|}
\hline
\multicolumn{1}{|c}{\textlst{FourMomentum} class API methods} & \\
\hline\hline
beta, gamma & \\
boostFromRestFrameOf & \\
boostToRestFrameOf & \\
boostVector & \\
dot & \\
eta, rapidity & \\
fromEtaPhiME & \\
fromPM & \\
fromPtEtaPhiM & \\
fromRoot & \\
mass, mass2 & \\
p, p2, PFlip, pVec & \\
phi, theta & \\
E, px, py, pz, pt  & \\
operator *=, /=, +=, -=, - & \\
dot & $\circ$ \\
operator *, +, -, / & $\circ$ \\
angle, costheta & $\circ$ \\
deltaR, deltaPhi, deltaEta & $\circ$ \\
epsilon & $\circ$ \\
boostToRestFrameOf & $\circ$ \\
\hline
\multicolumn{1}{|c}{\textlst{IOBuffer} class API methods} & \\
\hline\hline
kind, length, start & {\tiny{$\square$}}\\
load & \\
save & \\
init & \\
operator >>, << & $\circ$ \\
\hline
\multicolumn{1}{|c}{\textlst{IOBuffers} class API methods} & \\
\hline\hline
at & \\
operator [] & \\
front, back & \\
begin, end, rbegin, rend & \\
clear & \\
save & \\
operator << & $\circ$ \\
\hline
\multicolumn{1}{|c}{\textlst{Log} class API methods} & \\
\hline\hline
\multicolumn{2}{|l|}{\textlst{debug, error, info, log, trace, warn}} \\
getLevel & \\
getLog & \\
guard & \\
isActive & \\
resetWarningCounters & \\
setLevel, setLevels & \\
setShowLevel & \\
setShowLoggerName & \\
setShowTimestamp & \\
setUseColors & \\
setWarningMaxCount & \\
\hline
\end{tabular}
}}
\end{center}
\caption{API methods (or operators) in the user-facing classes of the \hammer library. A `$\circ$' denotes a method belonging to the \textlst{Hammer} namespace itself.
A `{\tiny{$\square$}}' indicates a member variable.}
\label{tab:api}
\end{table}
\FloatBarrier
\clearpage

\addtocontents{toc}{\protect\vspace*{10pt}}

\ifx\mcitethebibliography\mciteundefinedmacro
\PackageError{unsrtM.bst}{mciteplus.sty has not been loaded}
{This bibstyle requires the use of the mciteplus package.}\fi
\def\enquote#1{``#1''}
\expandafter\ifx\csname url\endcsname\relax
  \def\url#1{{\tt #1}}\fi
\expandafter\ifx\csname urlprefix\endcsname\relax\def\urlprefix{URL }\fi
\expandafter\ifx\csname eprint\endcsname\relax\def\eprint#1{\url{#1}}\fi

\end{document}